\documentclass[openacc]{rstransa}
\usepackage{hyperref}
\usepackage{float}
\usepackage[FIGTOPCAP]{subfigure}
\usepackage{ragged2e}
 \usepackage[numbers]{natbib}
\def\*#1{\mathbf{#1}}

\begin{document}

\title{Validation and application of the lattice Boltzmann \\ algorithm for a  turbulent immiscible Rayleigh-Taylor system}

\author{
H.S. Tavares$^{1}$, L. Biferale$^{2}$, M. Sbragaglia$^{2}$ and A.A. Mailybaev$^{1}$}

\address{$^{1}$Instituto de Matem\'atica Pura e Aplicada -- IMPA, Rio de Janeiro, Brazil  \\
$^{2}$Dept. Physics and INFN, University of Rome Tor Vergata, Italy}

\subject{fluid dynamics, computational physics}


\corres{Hugo S. Tavares\\
\email{hugoczpb@gmail.com}\\
Luca Biferale\\
\email{biferale@roma2.infn.it}\\
Mauro Sbragaglia\\
\email{sbragaglia@roma2.infn.it}\\
Alexei A. Mailybaev\\
\email{alexei@impa.br}
}

\begin{abstract}
We develop a multicomponent  lattice
Boltzmann (LB) model for the 2D Rayleigh--Taylor turbulence  with a  Shan-Chen pseudopotential  implemented on GPUs. In the immiscible case this method is able to accurately overcome the inherent numerical complexity caused by the complicated structure of the interface that appears in the fully developed turbulent regime. Accuracy of the LB model is tested both for early and late stages of instability. For the developed turbulent motion we analyze the balance between different terms describing variations of the kinetic and potential energies. Then, we analyze the role of  interface in the energy balance, and also the effects of the vorticity induced by the interface in the energy dissipation. Statistical properties are compared for miscible and immiscible flows. Our results can also be considered as a first validation step to extend the application of LB model to 3D immiscible Rayleigh-Taylor turbulence. 
\end{abstract}


\begin{fmtext}

\section{Introduction}

When a heavier fluid is suspended atop  a lighter fluid, the so-called Rayleigh--Taylor (RT) instability can develop, which eventually leads to a mixing layer with a turbulent motion called Rayleigh-Taylor turbulence. Physical experiments of the RT instability have been challenging due to the difficulty of sustaining an unstable density stratification necessary to set up the appropriate initial conditions for the instability~\cite{celani2009phase,ramaprabhu2004experimental,wilson1972excitatory,huang2007rayleigh,waddell2001experimental}. 

\end{fmtext}


\maketitle

Despite this limitation, considerable advances in numerical simulations of the Rayleigh-Taylor instability have been verified in the last decades, specially in the context of the systems with miscible fluids~\cite{chertkov2003phenomenology,biferale2010high,celani2006rayleigh,boffetta2017incompressible,biferale2018rayleigh,zhao2020energy,zhou2017rayleigh}. Fewer works have been dedicated to the immiscible 2D viscous case~\cite{celani2009phase,young2006surface,brackbill1992continuum,carles2006rayleigh,livescu2004compressibility,liang2019direct,huang2020late,hosseini2021lattice,liang2016lattice}, and most of them are devoted to early stages of the instability with little information about the state of developed turbulence. Only recently, the 2D immiscible RT instability was simulated for high Reynolds numbers~\cite{tavares2020immiscible,huang2020late,liang2019direct,liang2014phase,wang2019brief}. To the best of our knowledge, only~\cite{tavares2020immiscible} simulated a statistically homogeneous fully developed turbulent mixing layer compatible with classical phenomenological theories for immiscible RT turbulence~\cite{chertkov2003phenomenology,chertkov2005effects}. One of the reasons for this is the highly complicated pattern formed by the interfaces that appear in the immiscible case, originating high gradients and singularities in the solutions. This is a source of challenging numerical instabilities in many numerical methods for multicomponent fluids. To the best of our knowledge, only recently  the energy budget for miscible case was analyzed~\cite{zhao2020energy}, and no analysis of the immiscible case have been reported in the scientific literature.
	
	In this article, we investigate the immiscible RT turbulence using numerical simulations based on the multicomponent lattice-Boltzmann method with Shan-Chen pseudopotential model~\cite{kruger2017lattice,succi2018lattice,falcucci2007lattice}. This method is able to accurately overcome the inherent numerical complexity caused by the  structure of the interface that appears in the fully developed turbulent regime \cite{scarbolo2013unified,celani2009phase,young2006surface}. This method, similarly to other lattice Boltzmann methods, also admits parallel implementations in many situations, which is very important for statistical analyses that require a substantial number of simulations. We run several simulations of the RT turbulence (at least 10 simulations for each experiment presented in this article) in parallel on GPUs using CUDA with a computational grid of resolution $10.000 \times 5.000$. It is important to emphasize
	that some other diffuse interface methods
	may also be able to treat the same complex interfacial phenomena described in this article, see~\cite{chikatamarla2015entropic,leclaire2017generalized,geier2015conservative,liang2016lattice,de2019universal} for a recent discussion.
	
	In the present article, we perform a number of numerical tests justifying validity of the LB model for the RT instability. We analyse the energy budget in the RT flows with the emphasis on the effects of interface and dissipation, and comparing the results with the miscible RT system.
We verify that non-isotropic contributions to the stress tensor due to variations of order parameter are small in the miscible case. In the immiscible flow, these contributions grow in time following the increase of the interface length, but remain small compared to buoyancy and viscous contributions. Also, numerical anisotropy of the Shan-Chen force generates spurious currents~\cite{sbragaglia2007generalized,connington2012review} within thin diffuse interfaces, which do not affect most of our measurements but may interfere in the results for enstrophy. 
We note that, in the companion article~\cite{tavares2020immiscible}, we apply the proposed numerical scheme for the investigation of the long-time behavior of RT systems and verify a series of phenomenological predictions.    
	
	The paper is organized as follows. Section~\ref{section_2} describes details of the lattice Boltzmann model. In the section~\ref{section 3}, we study the evolution of the kinetic and potential energies of the systems, comparing miscible and immiscible systems side by side. In the subsection~\ref{subsection interface}, we analyze the influence of the interface in the energy balance for the immiscible Rayleigh-Taylor flows, by calculating the total energy of the interface and showing connections with different terms in the kinetic energy variations. Some differences due to the energy necessary to form the interfaces are expected~\cite{blanchette2009energy,desai2009dynamics,guo2015thermodynamically,liu2003phase}, which also lead to an extra generation of the vorticity~\cite{terrington2020generation,brons2014vorticity}. To investigate such possibility, we calculate numerically the energy of the interface and we compare with the energy flux due to the Korteweg stress tensor~\cite{anderson1998diffuse,guo2015thermodynamically}. The generation of vorticity at the interface is investigated by analyzing the evolution of the enstrophy in the subsection~\ref{subsection enstrophy}. We also analyze the existence of critical points in the transition to turbulent regimes in the immiscible case by 
	studying the evolution of the density profiles in the final part of the subsection~\ref{subsection interface}. We summarize the results in the Conclusion.

\section{Lattice Boltzmann model}
\label{section_2}

In this section, we describe the two-component lattice Boltzmann method for simulating immiscible and miscible Rayleigh-Taylor systems in Boussinesq approximation; we refer to~\cite{kruger2017lattice,succi2018lattice} and specially~\cite{tavares2020immiscible} for more details. In this method, spatial coordinates and time take values on the lattice with spacings $\Delta x$ and $\Delta t$, and the system is described by the interactions between two species of particles, A and B. Considering the so-called D2Q9 scheme, each particle is allowed to have nine velocities $\mathbf{c}_0,\ldots,\mathbf{c}_8$. These velocities are given by the vectors $(0,0)$, $(\pm c,0)$, $(0,\pm c)$ and $(\pm c,\pm c)$ with $c = \Delta x/\Delta t$, such that a particle either stays at the same or moves to a neighboring lattice point in a single time step. The system is described by the functions $f^s_i(\mathbf{x},t)$ determining the number of particles of component $s = A$ or $B$ and velocity $\mathbf{c}_i$ at a given point and time. The densities of each component and common velocity of the fluid are defined as
	\begin{equation}
	\rho_s(\mathbf{x},t)=\sum_{i} f_{is}(\mathbf{x},t),  \quad 
	\mathbf{u}(\mathbf{x},t) 
	=\dfrac{\sum_{s,i} f_i^s(\mathbf{x},t)\mathbf{c}_i/\tau_s}{\sum_{s}
	\rho_s(\mathbf{x},t)/\tau_s},
	\end{equation}
where $s = A, B$ and $i = 0,\ldots,8$. The total density is given by the sum $\rho = \rho_A+\rho_B$.

The evolution is governed by the lattice-Boltzmann equations with the Bhatnagar-Gross-Krook collision term~\cite{scarbolo2013unified} 
	\begin{equation}\label{Kinetic equation}
	f^s_i(\mathbf{x}+\mathbf{c}_i\Delta t,t+\Delta t)-f^s_i(\mathbf{x},t)
	=-\dfrac{1}{\tau_s}\left[ f^s_i(\mathbf{x},t)
	-f_i^{s(eq)}(\rho_s,\mathbf{u}+\tau_s\mathbf{F}_s/\rho_s )\right], 
	\end{equation}  
where $\tau_s$ and $\mathbf{F}_s$ are the relaxation time and the forcing term for component $s$, respectively. 
The right-hand side in (\ref{Kinetic equation}) describes the relaxation towards the local equilibrium distribution 
	\begin{equation}
	\label{Equilibrium distribution}
	f_i^{s(eq)}(\rho_s,\mathbf{u}')
	=\rho_s w_i\left(1+\dfrac{3\mathbf{c}_i\cdot \mathbf{u}'}{c^2}
	+\dfrac{9(\mathbf{c}_i\cdot \mathbf{u}')^2}{2 c^4}
	-\dfrac{3\mathbf{u}' \cdot \mathbf{u}'}{2c^2} \right),\quad
	\mathbf{u}' = \mathbf{u}+\frac{\tau_s\mathbf{F}_s}{\rho_s},
	\end{equation}
with the lattice sound speed $c_s =c/\sqrt{3}$ and constant weights $w_i$. These weights are expressed through velocity components $\mathbf{c}_i = (c_i^1,c_i^2)$ by the conditions
	\begin{equation}
	\sum_{i}w_ic_i^ac_i^b = c_s^2\delta_{ab},\quad
	\sum_{i}w_ic_i^ac_i^bc_i^cc^d_i = c_s^4\left(\delta_{ab}\delta_{cd}
	+\delta_{ad}\delta_{bc}+\delta_{ac}\delta_{bd}\right)
	\quad
	\textrm{for}\ \ a,b,c,d = 1,2,
	\end{equation}
where $\delta_{ab}$ is the Kronecker delta.

The forcing terms $\mathbf{F}_s = \mathbf{F}_{s}^{ff}+\mathbf{F}_{s}^{fb}+\mathbf{F}_{s}^{ext}$
contain three parts describing the fluid-fluid interaction, the fluid-boundary interaction and the external forces. The first is given by the Shan-Chen inter-molecular force as
	\begin{equation}\label{Shan-Chen forcinng term}
	\mathbf{F}_{s}^{ff}(\mathbf{x},t)
	= - G_{AB}\rho_s(\mathbf{x},t)
	\sum_{i}w_i\rho_{s'}(\mathbf{x}+\mathbf{c}_i\Delta t,t)\mathbf{c}_i,
	\end{equation}
with $s' = B$ and $s = A$ or vice versa. The choice of the pseudopotential function in \eqref{Shan-Chen forcinng term} is dictated by simplicity: this is the simplest choice allowing phase segregation when the two fluids interact via repulsive interactions. Of course, other choices are possible, with different impacts on the overall stability~\cite{kullmer2018transition}. Function (\ref{Shan-Chen forcinng term}) describes a system without self-interaction, where the coupling constant $G_{AB}$ controls the interaction between components $A$ and $B$. 
The interaction between fluid and boundary is given by
	\begin{equation}
	\mathbf{F}_s^{fb}
	=-G_{sb}\rho_s(\mathbf{x},t)\sum_{i}w_iS(\mathbf{x}+\mathbf{c}_i\Delta t)\mathbf{c}_i,
	\end{equation}
where $S(\mathbf{x})$ is the indicator equal to unity at boundary nodes and vanishing otherwise. The parameters $G_{Ab}$ and $G_{Bb}$ control interactions between fluid components and solid boundary; they relate to contact angles of fluids in the mixture. External forces are introduced as 
	\begin{equation}
	\label{eq3_Fex}
	\mathbf{F}^{ext}_A= -\rho_A \tilde{g} \,\mathbf{e}_y,
	\quad
	\mathbf{F}^{ext}_B= \rho_B \tilde{g}\,\mathbf{e}_y,
	\end{equation}
which yield the buoyancy forces in Boussinesq approximation, as we will see below.

Solutions of this model  approximate, in the continuum limit, the coupled Navier-Stokes and Cahn--Hillard equations~\cite{succi2018lattice,benzi2009mesoscopic,scarbolo2013unified} given by
\begin{eqnarray}
\rho \left( \dfrac{\partial \*u}{\partial t}+(\*u \cdot \nabla)\*u \right)&=&- \nabla \cdot \*P+\nabla \cdot \left[\eta \nabla \*u+ \eta\nabla \*u^T \right]-\phi \tilde{g}\mathbf{e}_y. \label{Momenum equation chapter 4} \\
\nabla \cdot \*u &=&0,\label{Incompressibility equation chapter 4} \\ 
\dfrac{\partial \phi}{\partial t}+\nabla \cdot (\phi \*u)&=& \nabla \cdot \left[M \nabla \mu \right] \label{CH equation chapter 4}
\end{eqnarray}
for the velocity field $\mathbf{u}(\mathbf{x},t)$, the total density $\rho(\mathbf{x},t)$ and the order parameter $\phi(\mathbf{x},t) = \rho_A-\rho_B$. Here $\eta$ is dynamic viscosity, $\mu$ is the chemical potential and $M$ is the mobility coefficient.
with $\*P$ being the momentum-flux tensor 
\begin{equation}\label{Total stress tensor decomposition}
\*P=p_b\*I+\*P^K+\*K^{(\tau)},
\end{equation}
where
\begin{eqnarray}
p_b&=&c_s^2\rho +\dfrac{G_{AB}c_s^2}{4}(\rho^2-\phi^2),\label{equation of state chapter 4}\\
\*P^K&=&\left[-\kappa\phi\Delta\phi-\dfrac{\kappa}{2}|\nabla \phi^2| \right] \*I +\kappa\nabla \phi \otimes \nabla \phi,\label{Korteweg stress tensor chapter 4} \\
\*K^{(\tau)}&=& c_s^4\dfrac{\rho_A \rho_B}{\rho}\left(\tau -\dfrac{1}{2} \right)^2\left(\dfrac{\nabla \rho_A}{\rho_A} -\dfrac{\nabla \rho_B}{\rho_B} \right)\otimes \left(\dfrac{\nabla \rho_A}{\rho_A} -\dfrac{\nabla \rho_B}{\rho_B} \right),
\label{Spurious contribution chapter 4}
\end{eqnarray}
with the coefficient $\kappa=c^4_s\dfrac{G_{AB}}{4}$ for the Shan-Chen method. The part $\*P^k$ is called the Korteweg stress tensor~\cite{anderson1998diffuse,joseph1996non,joseph2010fluid}, $p_b$  denotes the pressure in the bulk regions and $\*K^{(\tau)}$ is an extra spurious $\tau$-dependent contribution, which is small for $\tau$ close to $1/2$ and for small variations of the total densities of the system. Explicit expressions for the chemical potential $\mu$ and the mobility coefficient can be found in~\cite{scarbolo2013unified}, such expressions are not used in this article. Notice that Equations \eqref{Incompressibility equation chapter 4}-\eqref{CH equation chapter 4} are obtained via the Chapman-Enskog expansion~\cite{kruger2017lattice,succi2018lattice} assuming that fields vary smoothly in time and space. In such limit the forcing implementation  \eqref{Equilibrium distribution} introduces spurious non-Galilean invariant terms. In practice, especially when dealing with numerical simulations developing sharp gradients in space and time, hydrodynamic consistency checks are needed (see Fig.~\ref{Consistency checks and components of the  kinetic energy variation}).

We choose $\Delta x = \Delta t =1$ (considered as lattice-Boltzmann units) in the rectangular
domain of horizontal size $L_x = 10^4$ and vertical size $L_y = L_x/2$. Periodic boundary conditions are assumed in the horizontal direction with the solid bottom and top boundaries. The bounce-back relation \cite{succi2018lattice,li2020multiscale} is used for the distribution function $f^s_i(\mathbf{x},t)$ at the solid boundaries for modeling the no-slip condition. The  relaxation time $\tau = 0.53$ is chosen for both components, providing the kinetic viscosity $\nu = c_s^2(\tau-1/2) = 0.01$. For small fluid velocities (small lattice Mach numbers) $|\mathbf{u}|\ll c_s$, the flow can be assumed incompressible. We consider pure densities of both fluid components equal to $1.10$ and the gravity parameter $\tilde{g} = 9\cdot 10^{-6}$.  This gravity value is sufficient to overcome the effects of surface tension without generating significant fluctuations in total density. Since changes of the total density due to pressure variations and mixing are small, we approximate $\rho(\mathbf{x},t) \approx \rho_0$ by a constant.

The coupling constant $G_{AB}$ has a critical value with the immiscible (two phase) flow for stronger couplings and miscible (single phase) flow for weaker couplings. For our immiscible and miscible models we select $G_{AB}=0.1381$ and $G_{AB}=0.0805$, respectively.  In the interactions with the boundaries we use neutral wetting, i.e., $G_{Ab}=G_{Bb}=0$, to minimize the influence of the boundaries in the simulations. In the immiscible model, two phases are separated by a diffuse interface having a width of approximately $l_{int} \sim 3$ grid nodes. This model approximates the Boussinesq system for the immiscible RT systems~\cite{tavares2020immiscible} considered at scales much larger than $l_{int}$ with the surface tension $\sigma = 0.0059$ obtained from pressure measurements for large bubbles. Similarly, one recovers the miscible Boussinesq approximation for the miscible RT systems~\cite{tavares2020immiscible} in the continuous limit for small gradients of the order parameter. The diffusion coefficient can be estimated roughly as $\mathcal{D} \simeq c_s^2\left[ \left(\tau -1/2\right) -\rho  \tau G_{AB}/2 \right] = 0.002$~\cite{shan1996diffusion,benzi2009mesoscopic}. For simulations in this paper, we initialize the flow by using an equilibrium immiscible configuration and adding a small random (white-noise) deformation to the interface with an amplitude of 4 grid points. In this equilibrium configuration, the first phase consists primarily of component $A$ with about $9\%$ of component $B$, and vise versa for the second phase.

The simulations are implemented on GPUs of the model NVIDIA Tesla V100 PCIe 32 GB. This helps to accumulate suitable statistics with a reasonable amount of time.
Specifically, we consider ensembles with at least 10 simulations. Every simulation takes around 10 hours to perform 90.000 time steps, which is enough to obtain fully developed mixing layers for miscible and immiscible RT systems. The same simulation using sequential codes in CPUs takes a few days. For further quantitative descriptions on the GPU codes in use we refer to~\cite{bernaschi2009graphics,bernaschi2017gpu,pelusi2019impact}.   In \cite{bernaschi2009graphics} some performance measurements for soft flows under periodic shear indicate a GPU/
CPU speed up ranging from 2 to 12 for grids from $128 \times 128$ to $1024 \times 1024$. A bigger difference is expected for bigger grids.

\section{Results and discussion}
\label{section 3}

\begin{figure}[H]
	\centering
	\includegraphics[width=1.0\textwidth]{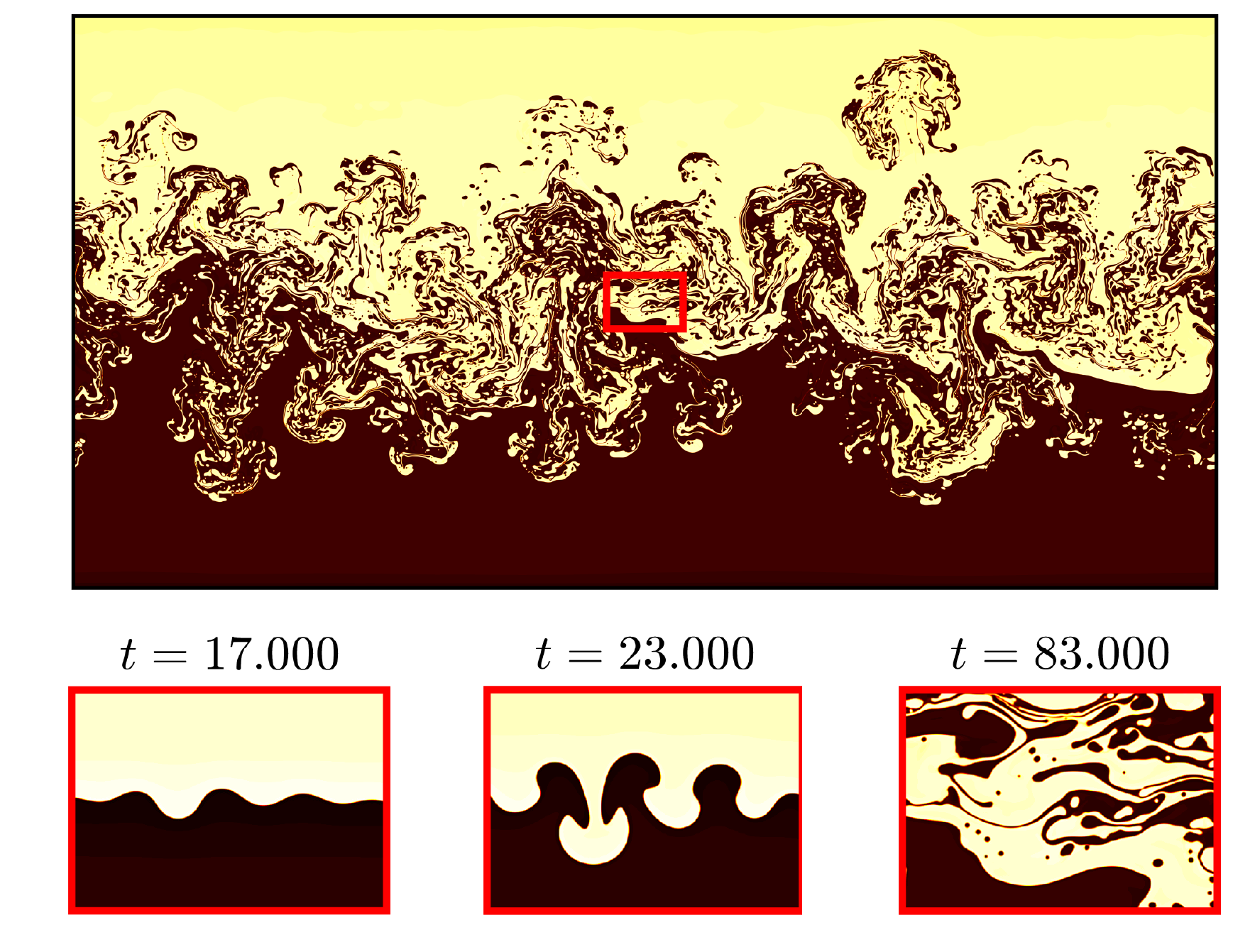}
	\includegraphics[width=1.0\textwidth]{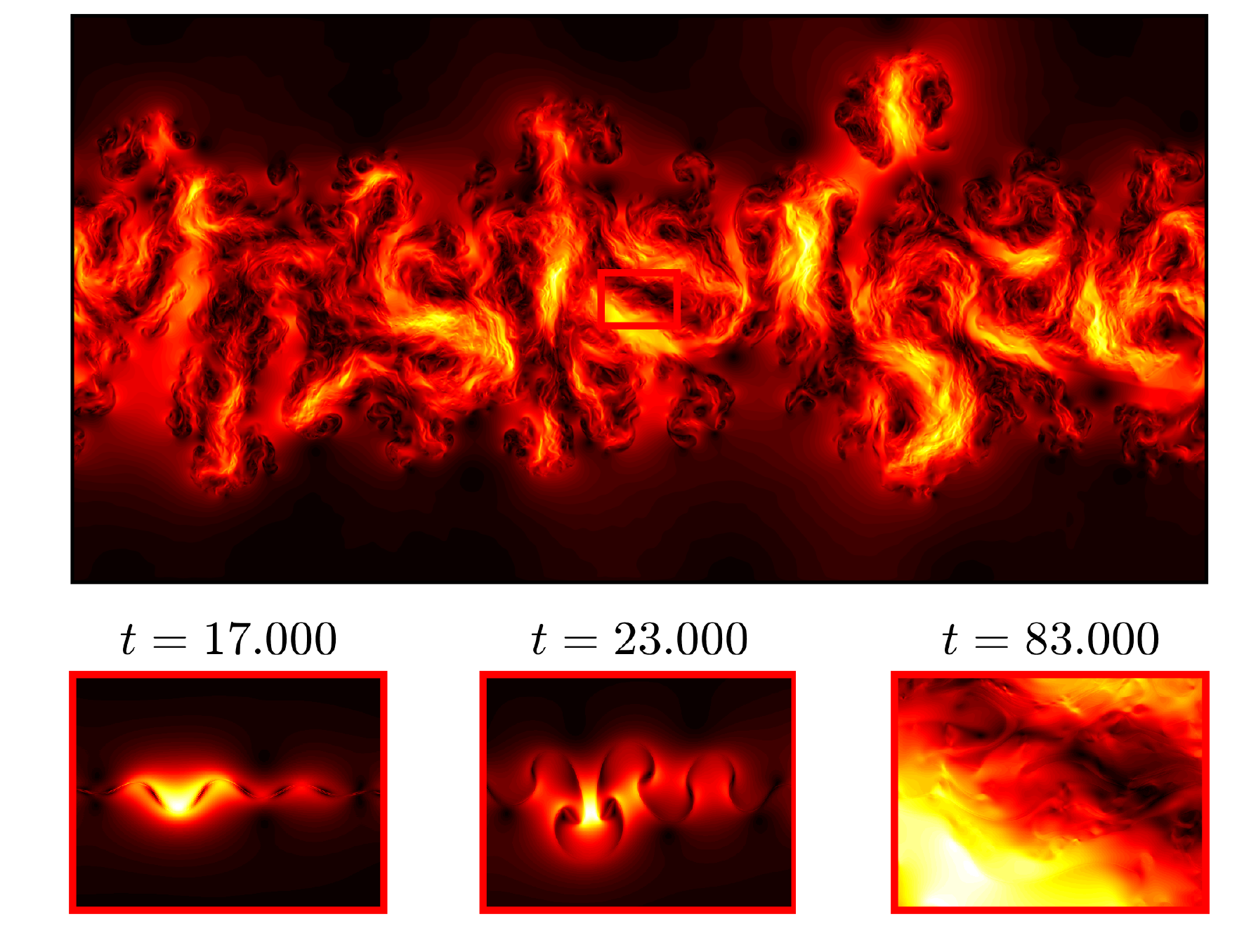}
	\caption{Mixing layer of the immiscible Rayleigh-Taylor turbulence. Upper set of pictures describes the density field, where the yellow color represents a heavier phase and the brown color corresponds to a lighter phase. The lower set of pictures describes the absolute value of velocity, with lighter colors corresponding to larger velocities.
	Small pictures show zooms into the small region (marked in the center of the main panel) for three different times. These times correspond to the initial linear growth, formation of nonlinear mushroom-like structures at intermediate times, and fully developed turbulent mixing at larger times. Simulations are performed on the grids $10.000 
		\times 5.000$ in lattice-Boltzmann units.}
	\label{fig1_novo}	
\end{figure}

\begin{figure}[H]
	\centering
	\includegraphics[width=1.0\textwidth]{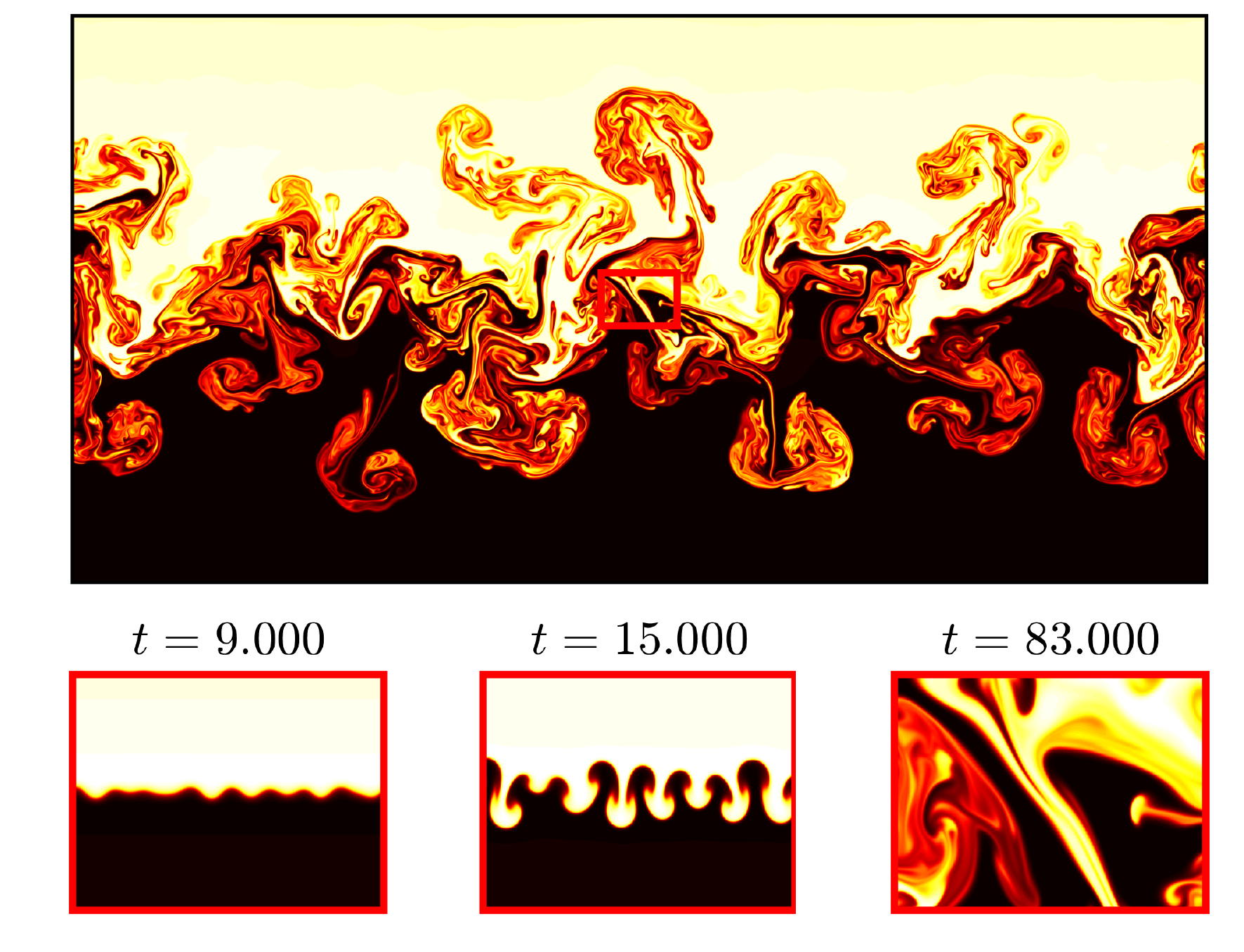}
	\includegraphics[width=1.0\textwidth]{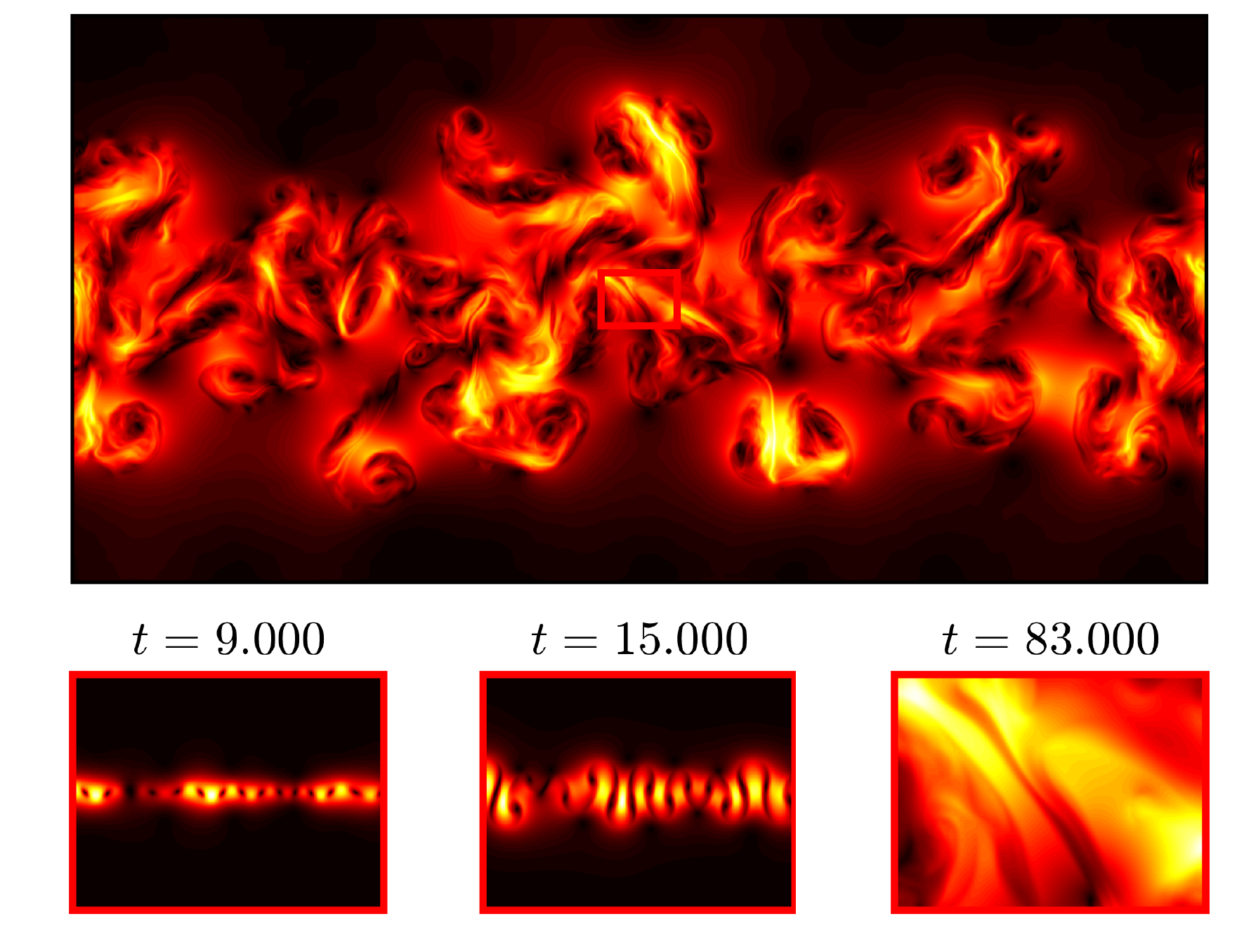}
	\caption{Same plots as in Fig.~\ref{fig1_novo} but now for the miscible Rayleigh-Taylor turbulence.}
	\label{fig2_novo}	
\end{figure}

We first develop the equation for the kinetic and potential energies of the system. Then, we study the  evolution of such energies for the immiscible and miscible RT system using numerical data from the Shan-Chen multicomponent method.  The results for the density and velocity fields are shown in the Figures~\ref{fig1_novo} and~\ref{fig2_novo} for the immiscible and miscible cases, respectively. In this figures, we can observe that after an initial linear growth the perturbations develop into nonlinear mushroom-like structures evolving further to the fully developed turbulent mixing layer. The important difference between the immiscible and miscible cases can be seen at small scales. The immiscible Rayleigh-Taylor turbulence leads to the formation of emulsion-like state with a multitude of small bubbles. The miscible Rayleigh-Taylor turbulence develops sharp gradients leading to enhanced diffusion at small scales.

We analyse the potential and kinetic energies of the RT systems defined as~\cite{celani2009phase}
\begin{equation}
    E_p=\tilde{g} \left\langle y \phi \right\rangle, \quad E_k=\left\langle \rho \dfrac{|\*u|^2}{2} \right\rangle, \label{Kinetic energy balance}
\end{equation}
where the averages are calculated in a subdomain  obtained by cutting the 10 rows closest to the top and the 10 rows closest to the bottom of the original computational domain, to avoid complications with the boundaries. This means that the subdomain has the size $D=10.000\times 4.980$. The time derivatives and differential operators are calculated numerically by using centered finite difference schemes.

 The variations of potential and kinetic energies, obtained from (\ref{Momenum equation chapter 4})--(\ref{CH equation chapter 4}), are given by
\begin{eqnarray}
\partial_t E_p&=& \left\langle \*u \cdot \phi \tilde{g}\*e_y\right\rangle,  \\[2pt]
\partial_t E_k&=&- \left\langle \*u  \cdot (\nabla \cdot \*P)\right\rangle +\left\langle  \*u \cdot \left(\nabla \cdot \left( \eta \nabla \*u+\eta \nabla \*u^T\right)\right)\right\rangle   - \left\langle \*u \cdot \phi \tilde{g}\*e_y \right\rangle. \label{Kinetic energy balance}
\end{eqnarray}
We can see three different contributions to the kinetic energy variation:
\begin{eqnarray}\label{Contributions for kinetic energy variations}
\begin{aligned}
&\textrm{Contribution of the momentum-flux tensor:} \quad -\left\langle \*u  \cdot (\nabla \cdot \*P)\right\rangle.\label{Contribution from the pressure tensor}\\
&\textrm{Contribution of the viscous term:}\quad \left\langle  \*u \cdot \left( \nabla \cdot \left( \eta \nabla \*u+\eta \nabla \*u^T \right)\right)\right\rangle.\label{Contribution from the viscous term}\\
&\textrm{Contribution of the buoyancy term:}\quad
\left\langle \*u \cdot \phi \tilde{g} \*e_y \right\rangle.\label{Contribution of the buoyancy term}
\end{aligned}
\end{eqnarray}

\begin{figure}[!h]
    \subfigure[]{\includegraphics[width=.44\linewidth]{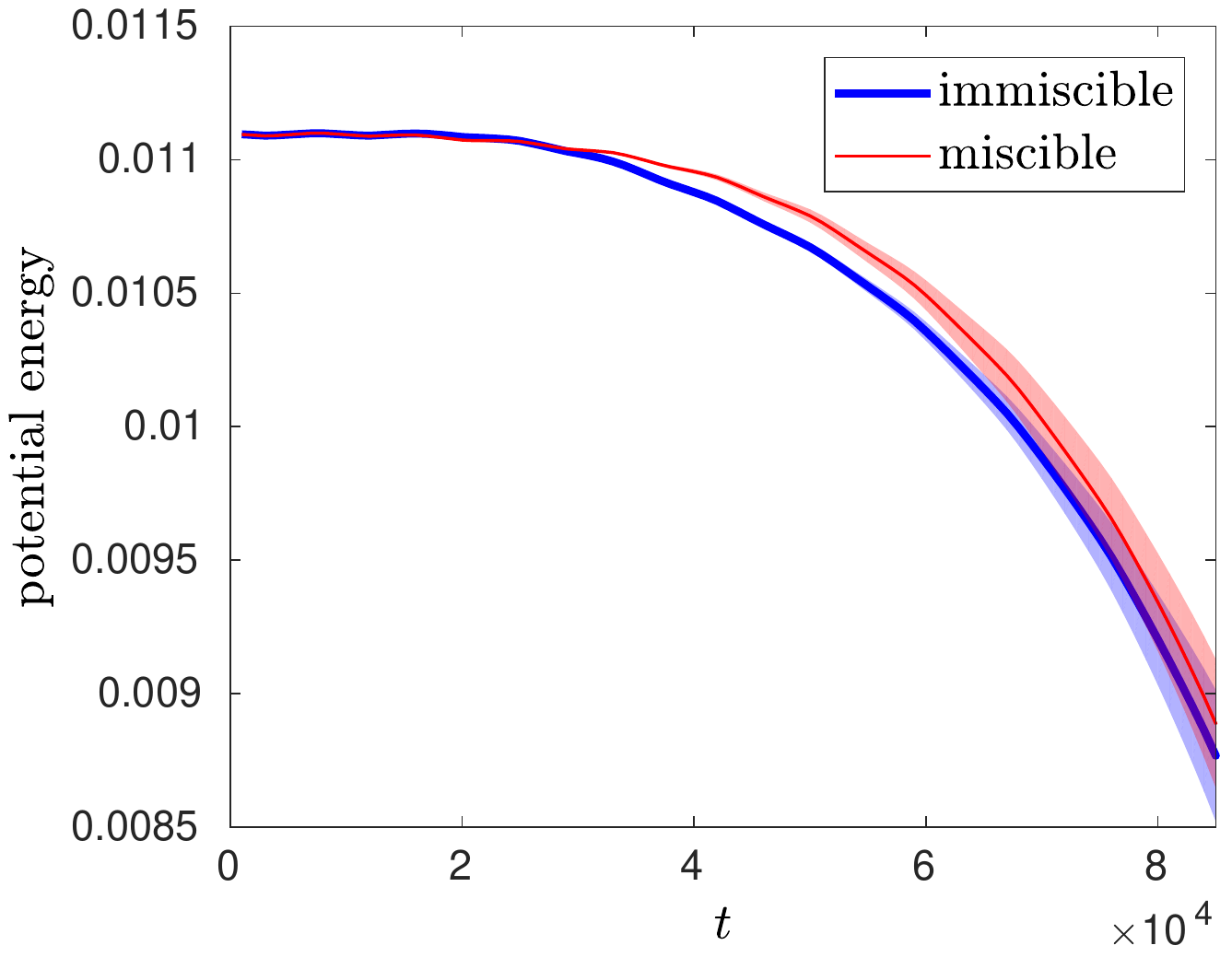}}\qquad
	\subfigure[]{\includegraphics[width=.44\linewidth]{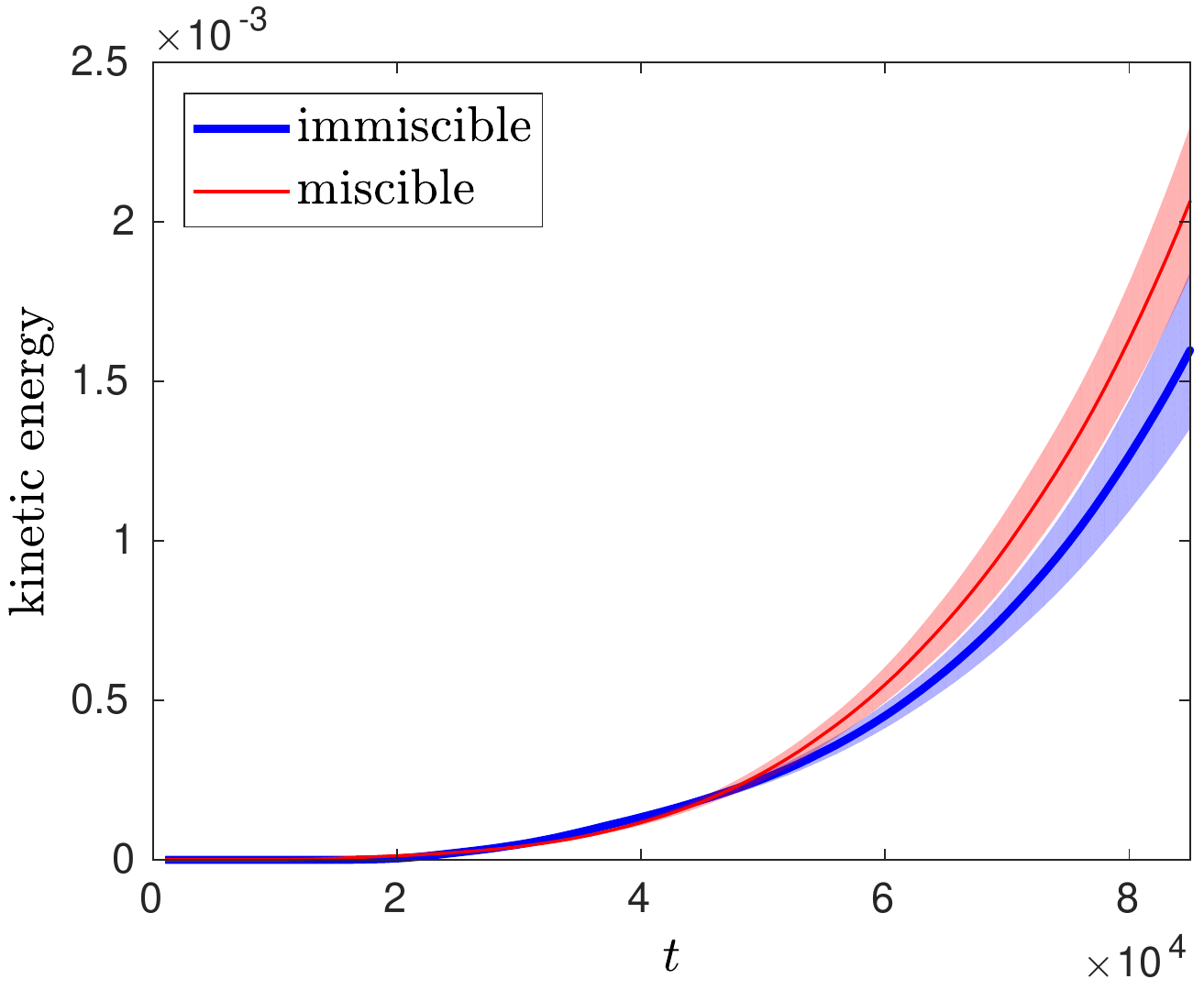}}\\
    \subfigure[]{\includegraphics[width=.44\linewidth]{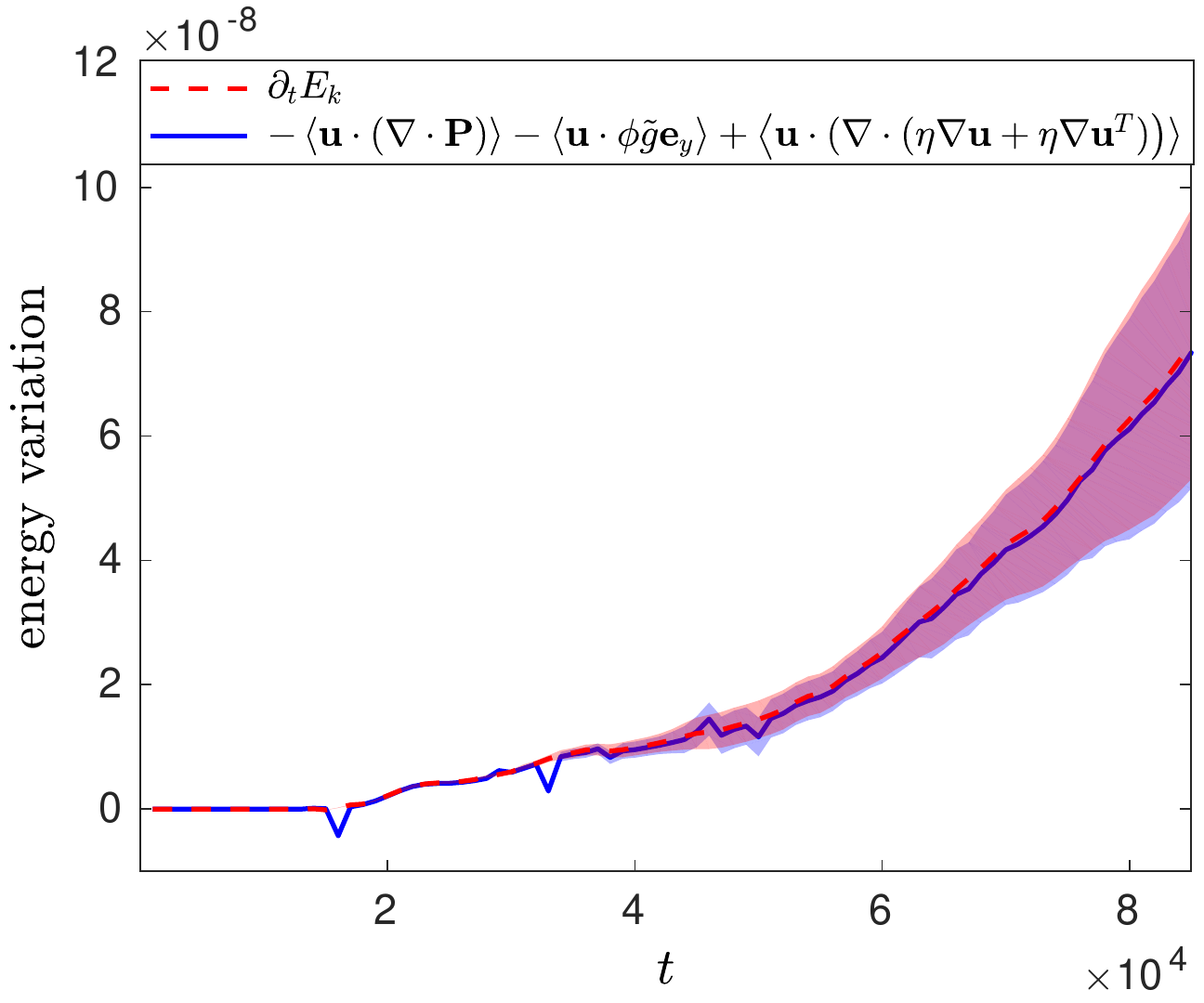}}\qquad
    \subfigure[]{\includegraphics[width=.44\linewidth]{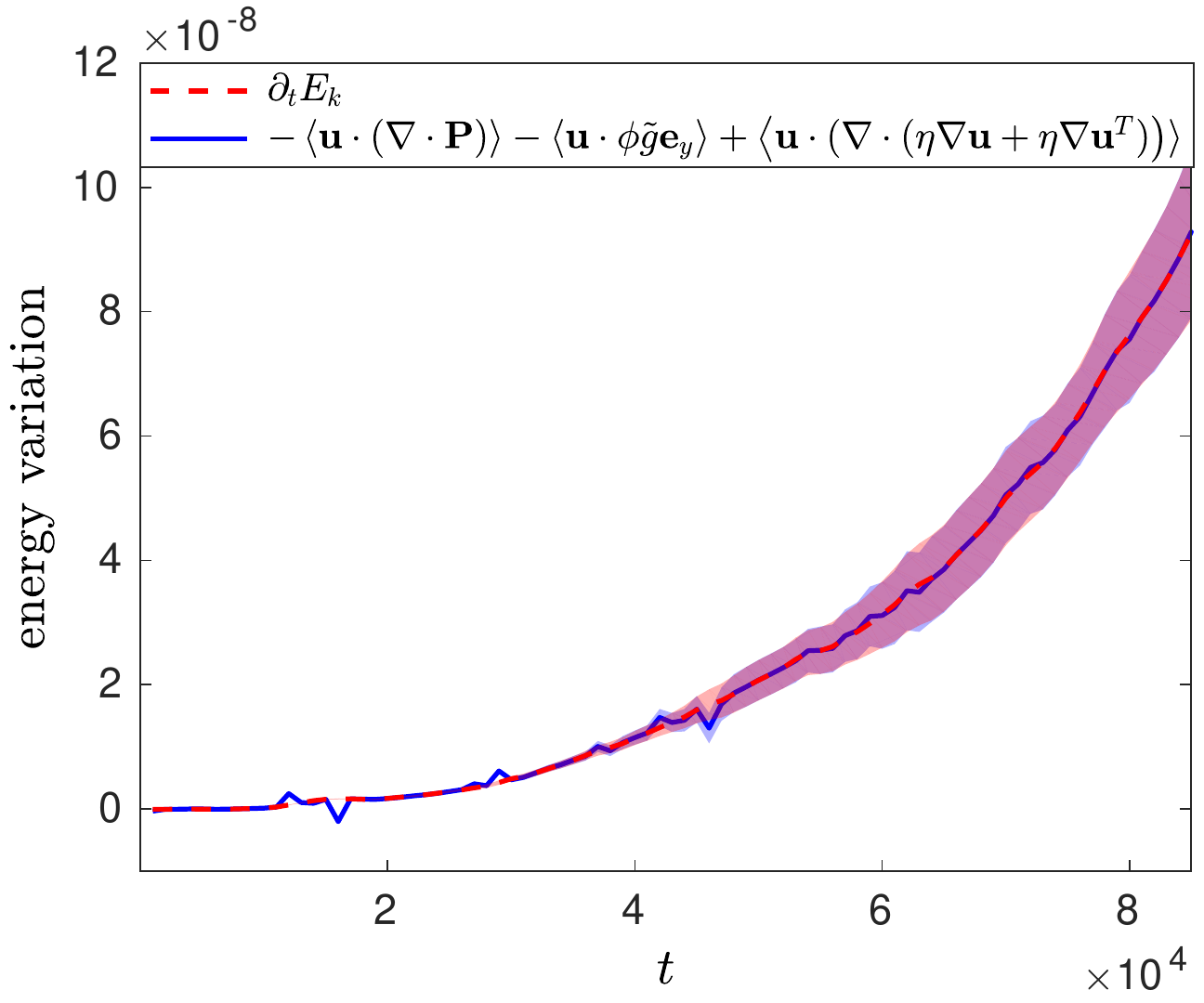}}\\
	\subfigure[]{\includegraphics[width=.44\linewidth]{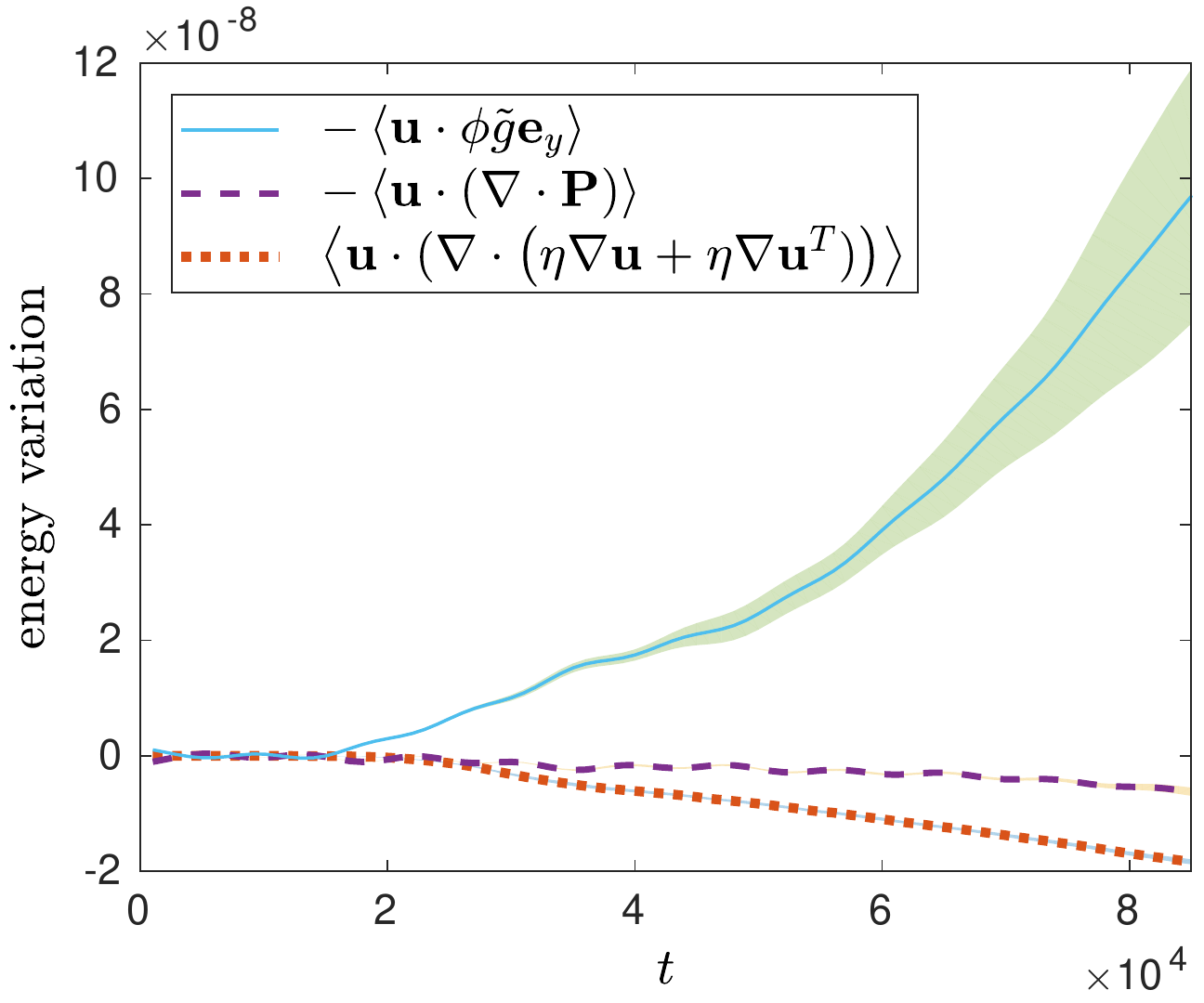}}\qquad
	\subfigure[]{\includegraphics[width=.44\linewidth]{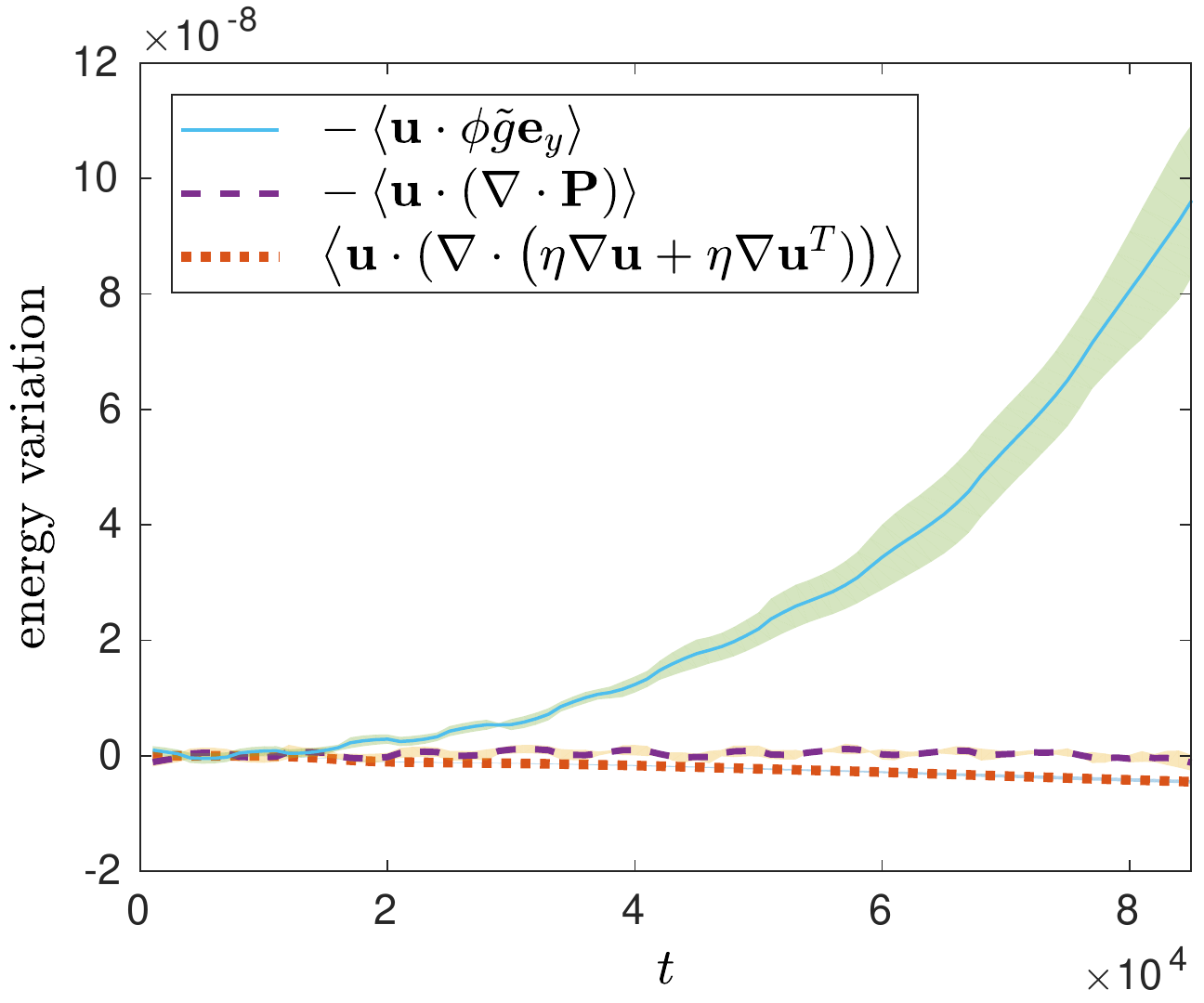}}
 \caption{(a) Evolution of the potential energy $E_p$ for miscible and immiscible flows. (b) Evolution of the kinetic energy $E_k$ for miscible and immiscible flows. Verification of the kinetic energy balance for immiscible (c) and miscible (d) flows. Components of the kinetic energy variation (\ref{Kinetic energy balance}) for immiscible (e) and  miscible (f) flows. The shaded regions indicate standard deviations.}
\label{Consistency checks and components of the  kinetic energy variation}
\end{figure}

The potential and kinetic energies are shown in Figs.~\ref{Consistency checks and components of the  kinetic energy variation}(a, b), where we compare the averages and standard deviations using 10 simulations obtained for immiscible and miscible flows. The kinetic energy grows faster in the miscible case. In Figs.~\ref{Consistency checks and components of the  kinetic energy variation}(c, d), we verify the balance (\ref{Kinetic energy balance}) for immiscible and miscible systems. These graphs demonstrate a good agreement confirming that the solutions for the order parameter $\phi$ and the velocity field $\*u$ obtained by the Shan-Chen multicomponent method  satisfy accurately the coupled Navier-Stokes and Cahn--Hillard system (\ref{Momenum equation chapter 4})--(\ref{CH equation chapter 4}).

In Figs.~\ref{Consistency checks and components of the  kinetic energy variation}(e, f) we present separately the contribution of each term in (\ref{Contribution from the viscous term}) to the kinetic energy balance. One can see that the difference between the immiscible and miscible flows in the growth of kinetic energy in Fig.~\ref{Consistency checks and components of the  kinetic energy variation}(b) is associated mainly with the terms $\nabla \cdot \*P$ and $ \nabla \cdot \left( \eta \nabla \*u+\eta \nabla \*u^T\right)$. Here, the first term describes the contribution of the momentum flux tensor responsible for the effects of the surface tension in the immiscible Rayleigh-Taylor system. It is small in comparison with the contribution of the buoyancy term. This difference has important implications in the analysis of the long time behavior of Rayleigh-Taylor systems~\cite{tavares2020immiscible}.

In Fig.~\ref{Components of the pressure tensor}, we study the decomposition (\ref{Total stress tensor decomposition})--(\ref{Spurious contribution chapter 4}) by analyzing the expression 
 \begin{equation}\label{Three parts of the pressure tensor}
 \left\langle \*u  \cdot (\nabla \cdot \*P)\right\rangle=\left\langle \*u  \cdot \nabla p_b\right\rangle+\left\langle \*u  \cdot (\nabla \cdot \*P^K)\right\rangle+\left\langle \*u  \cdot (\nabla \cdot \*K^{(\tau)})\right\rangle.
 \end{equation}
 Fig.~\ref{Components of the pressure tensor}(a) shows that the contribution of the spurious term $\*K^{(\tau)}$ is negligible in comparison with the other terms, therefore, it does not generate a significant impact in the measurements of the energy flux.  The energy flux term $\left\langle \*u  \cdot \nabla p_b\right\rangle$ is of the same order for miscible and immiscible flows, and its oscillatory aspect is essentially caused by density fluctuations due to the initialization process of the lattice-Boltzmann 
 algorithm~\cite{kruger2017lattice}. A significant difference between miscible and immiscible flows is related to the Korteweg stress tensor $\*P^K$ presented in Fig.~\ref{Components of the pressure tensor}(b). Below we explain that this difference represents the portion of kinetic energy which is converted into the energy of interface.

\subsection{Energy of the interface}
\label{subsection interface}

In this subsection, we show that the energy flux $	\left\langle (\nabla \cdot \*P^{K})\cdot \*u \right\rangle$ due to the Korteweg stress tensor is directly connected with the variation of the total energy of the interface defined as the product between the total length $\mathcal{L}$ of the interface and the surface tension $\gamma$.

Points of the moving interface $\Gamma(t)$ for an immiscible binary mixture are commonly given by the equation $\phi(\*x,t)=0$. 

\begin{figure}[t]
 \subfigure[]{\includegraphics[width=.44\linewidth]{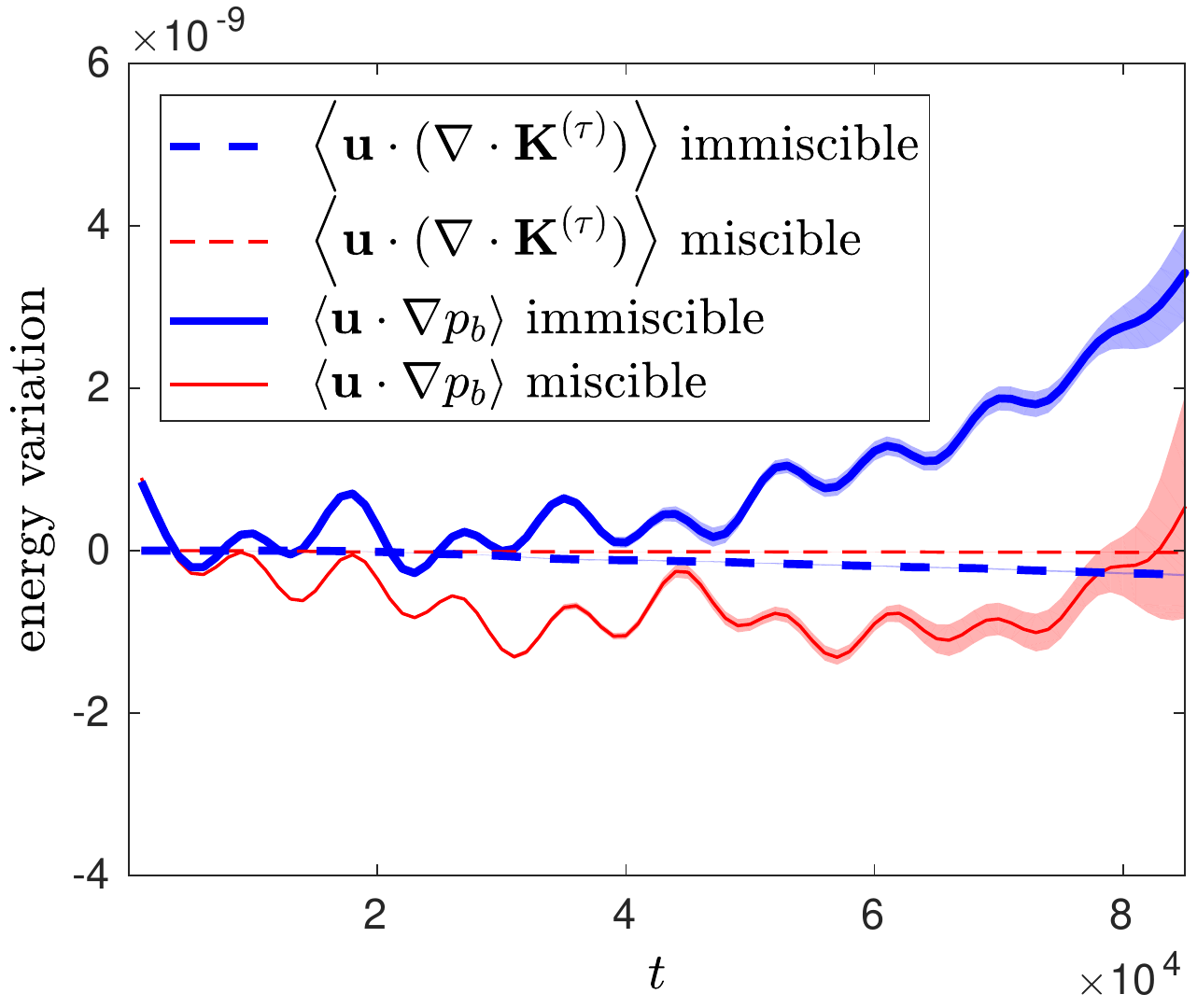}}\qquad
    \subfigure[]{\includegraphics[width=.44\linewidth]{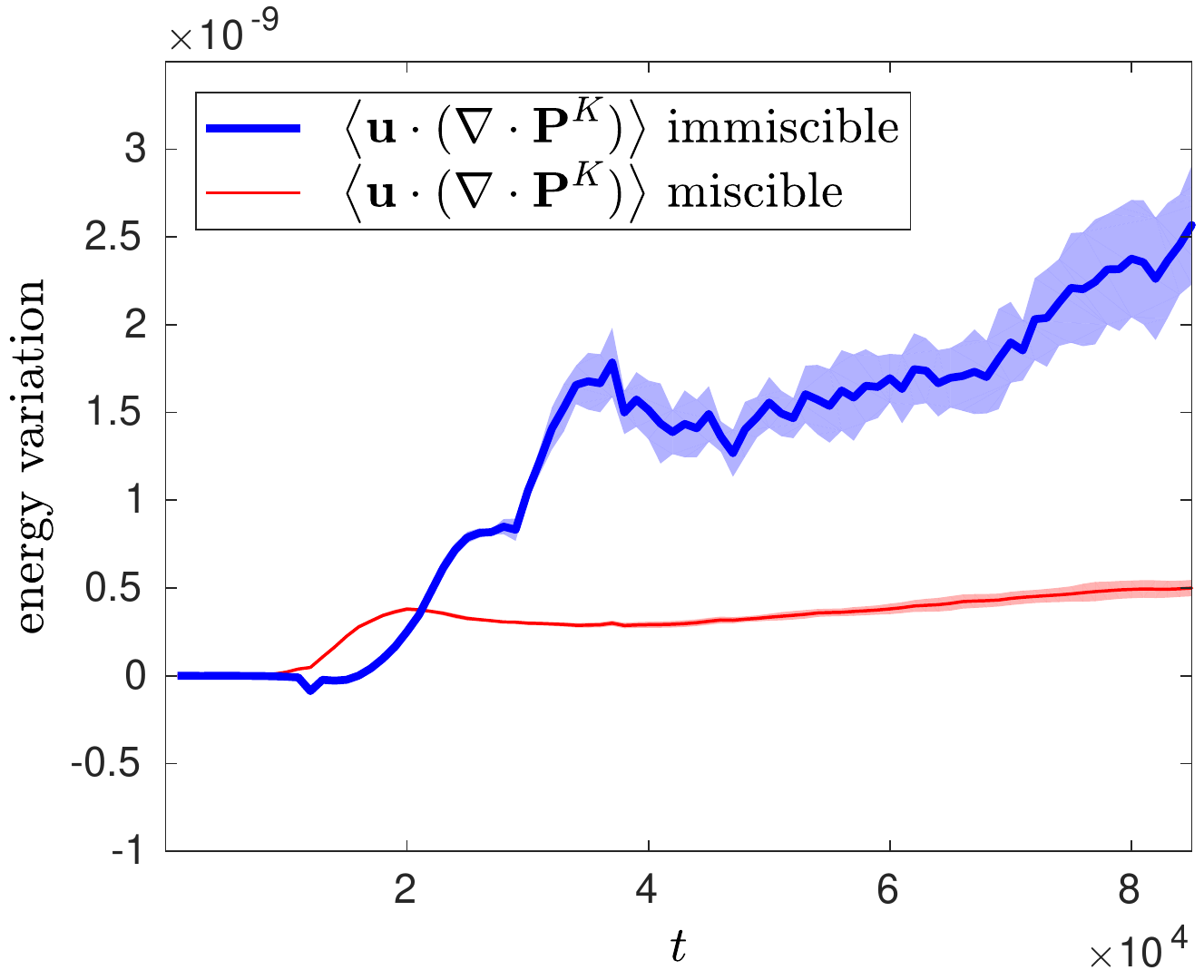}}\\
	\subfigure[]{\includegraphics[width=.44\linewidth]{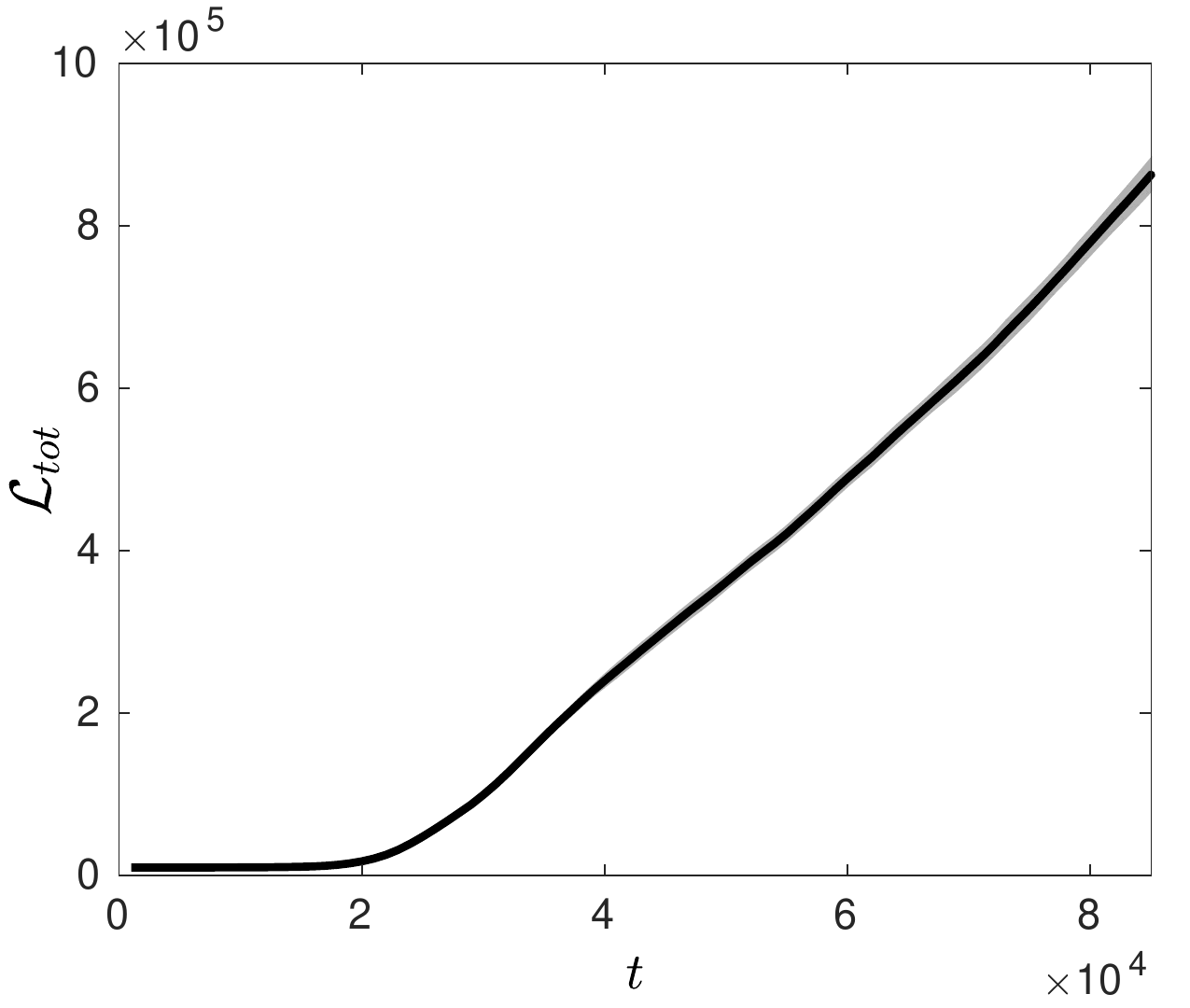}}\qquad
	\subfigure[]{\includegraphics[width=.44\linewidth]{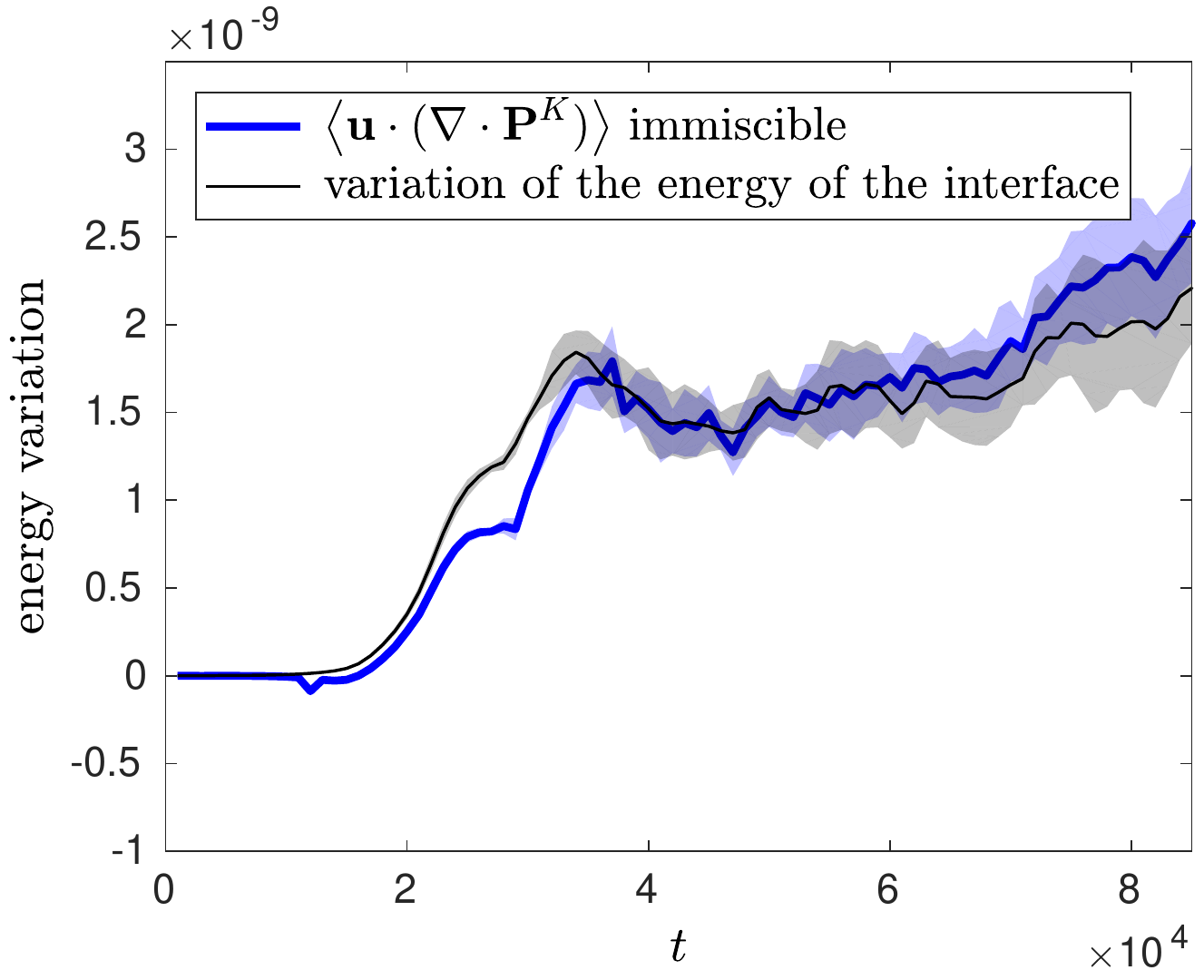}}
 \caption{(a) Comparison between the energy flux due to the bulk pressure and spurious term. (b) Comparison between the energy flux due to the Korteweg stresses. (c) Evolution of the total length. (d) Comparison between the energy flux due to the Korteweg stresses and the right hand side of (\ref{Korteweg stress and total length}) indicated as the variation of the energy of the interface. The shaded regions indicate standard deviations.}
\label{Components of the pressure tensor}
\end{figure}

 When the curvature
radius is large with respect to the interface thickness, we have the following relations~\cite{tavares2021,liu2003phase,chella1996mixing,scarbolo2015coalescence,anderson1998diffuse}
\begin{equation}\label{Curvature and stress tensor}
\begin{array}{ll}
\displaystyle
\int_{\Omega}(\nabla \cdot \*P^K)\cdot\*u~d\*x 
&\displaystyle
=- \kappa\int_{\Omega}\phi \nabla (\Delta \phi)  \cdot \*u~d\*x 
\simeq -\kappa\int_{\Omega}   |\nabla \phi|^2 K \left( \dfrac{\nabla \phi}{|\nabla \phi|} \cdot \*u \right)~d\*x \\[12pt]
&
\displaystyle
\simeq -\int_{\Gamma(t)} \gamma K (\*n \cdot\*u) ds.
\end{array}
\end{equation}
where $\*n=\frac{\nabla \phi}{|\nabla \phi|}$ and is the normal field on $\Gamma(t)$, $K=-\nabla \cdot \*n$~\cite{goldman2005curvature} is the scalar curvature and $\*P^K$ is given by (\ref{Korteweg stress tensor chapter 4}).
Using the {\it total length equation} described in~\cite{sevcovic2001evolution}, it can be shown that the left-hand-side in (\ref{Curvature and stress tensor}) can be rewritten in a such way that
\begin{equation}\label{Korteweg stress and total length}
\left\langle (\nabla \cdot \*P^{K})\cdot \*u \right\rangle\simeq \dfrac{1}{D}\dfrac{d(\gamma \mathcal{L}_{tot})}{dt},
\end{equation}	
where $\mathcal{L}_{tot}$ is the total length of the interface between two phases.
This formula implies that the energy flux due to the Korteweg stress tensor (\ref{Korteweg stress tensor chapter 4}) corresponds directly to the variation of the energy of the interface. 
We studied the flux $\left\langle (\nabla \cdot \*P^{K})\cdot \*u \right\rangle$ for miscible and immiscible RT flows in  Figs.~\ref{Components of the pressure tensor}(b, d). In Fig.~\ref{Components of the pressure tensor}(d) we verify the relation (\ref{Korteweg stress and total length}) showing that the difference in miscible and immiscible flows is associated with the appearance of the interface. The energy of the interface is calculated as the product of the surface tension $\gamma$ and the total length of the interface $\mathcal{L}_{tot}$. The latter is calculated by the Cauchy-Crofton formula~\cite{do2016differential,legland2007computation} with the result shown in Fig.~\ref{Components of the pressure tensor}(c). A small difference between the two curves in Fig.~\ref{Components of the pressure tensor}(d) can be attributed to the diffuse interface assumption in the lattice-Boltzmann algorithm and to high values of curvatures in the late stages of the immiscible Rayleigh-Taylor turbulence. 

\begin{figure}[t]
	\centering
	\subfigure{\includegraphics[scale=0.56]{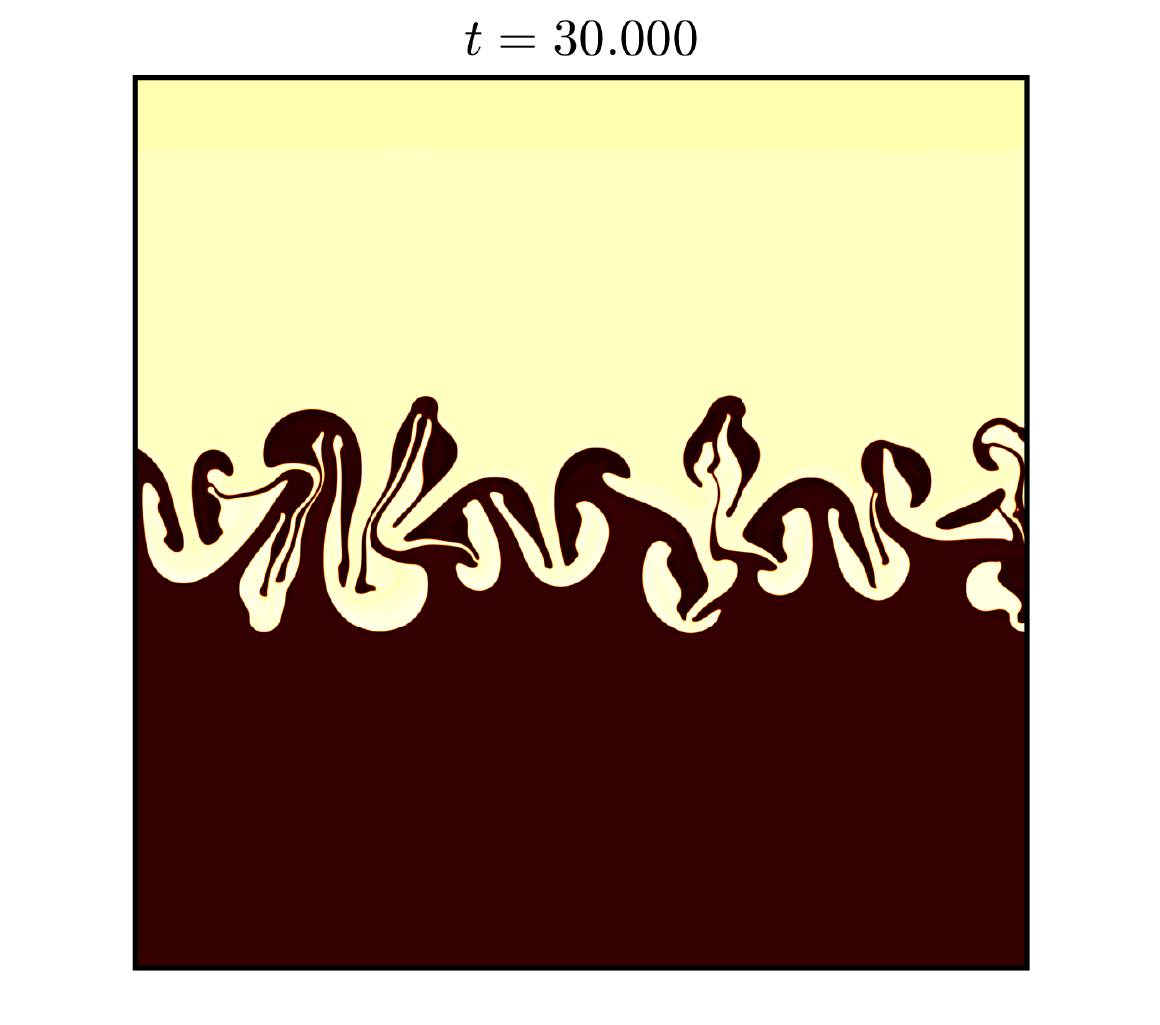}}
	\subfigure{\includegraphics[scale=0.56]{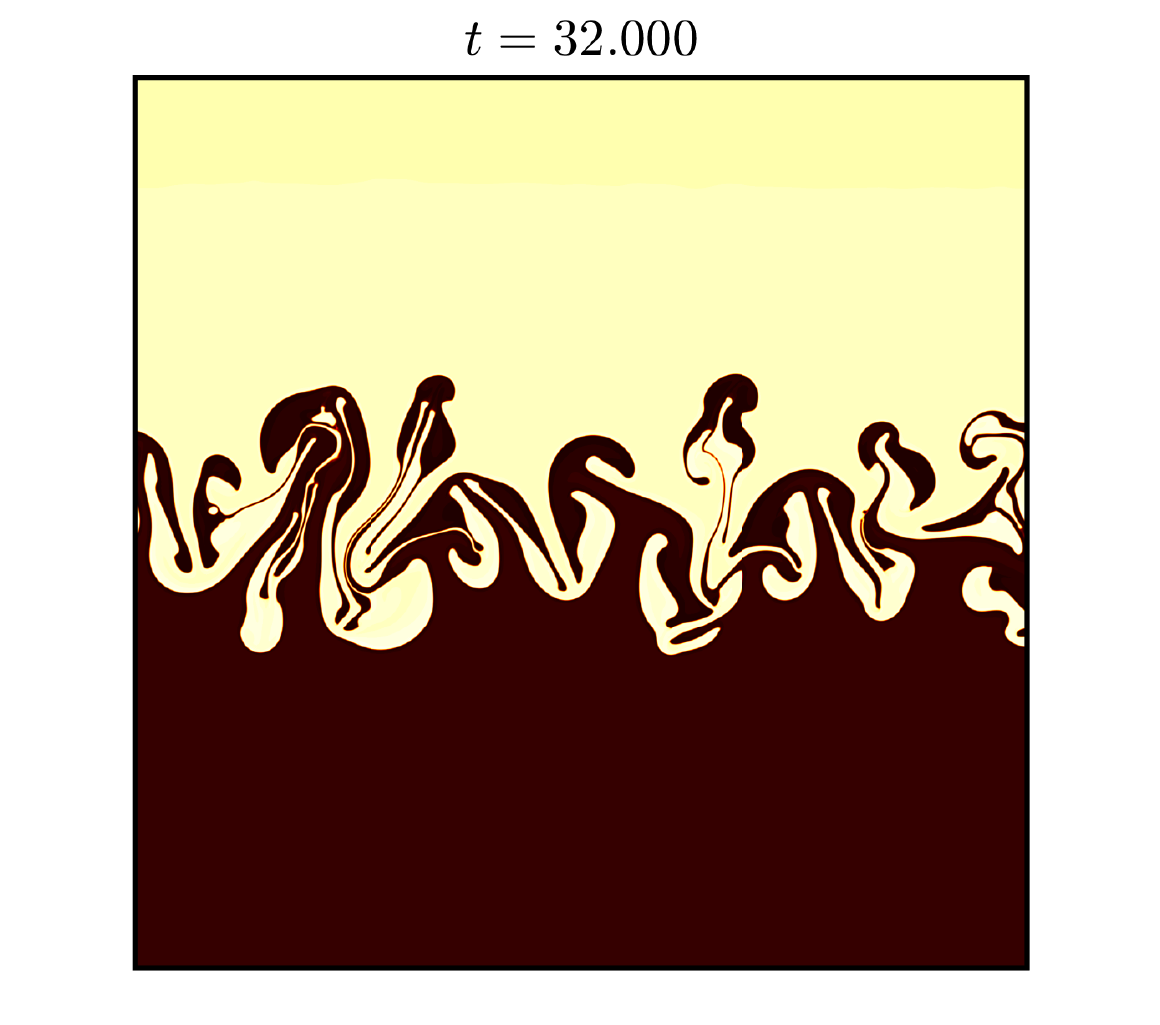}}\\
	\subfigure{\includegraphics[scale=0.56]{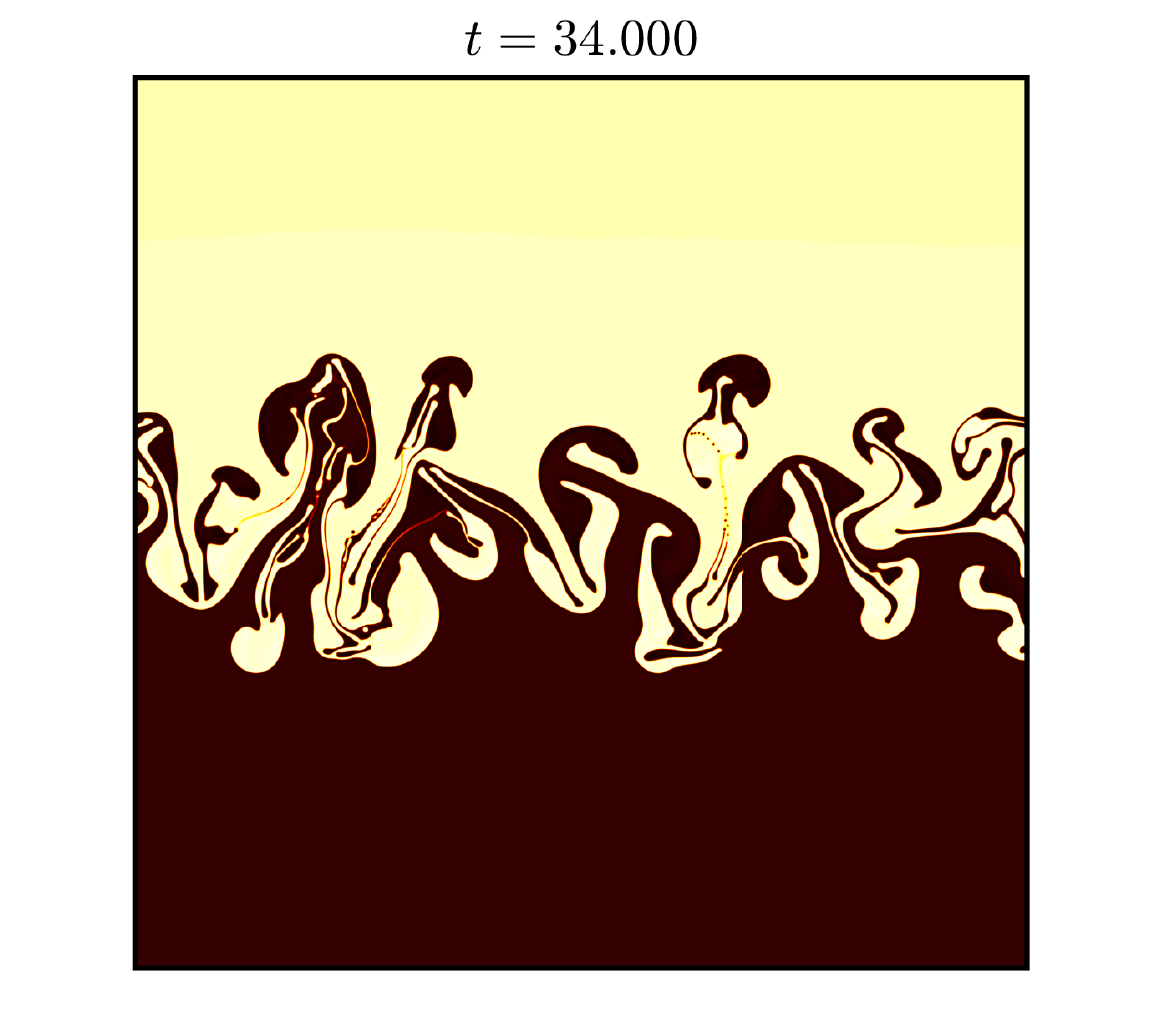}}
	\subfigure{\includegraphics[scale=0.56]{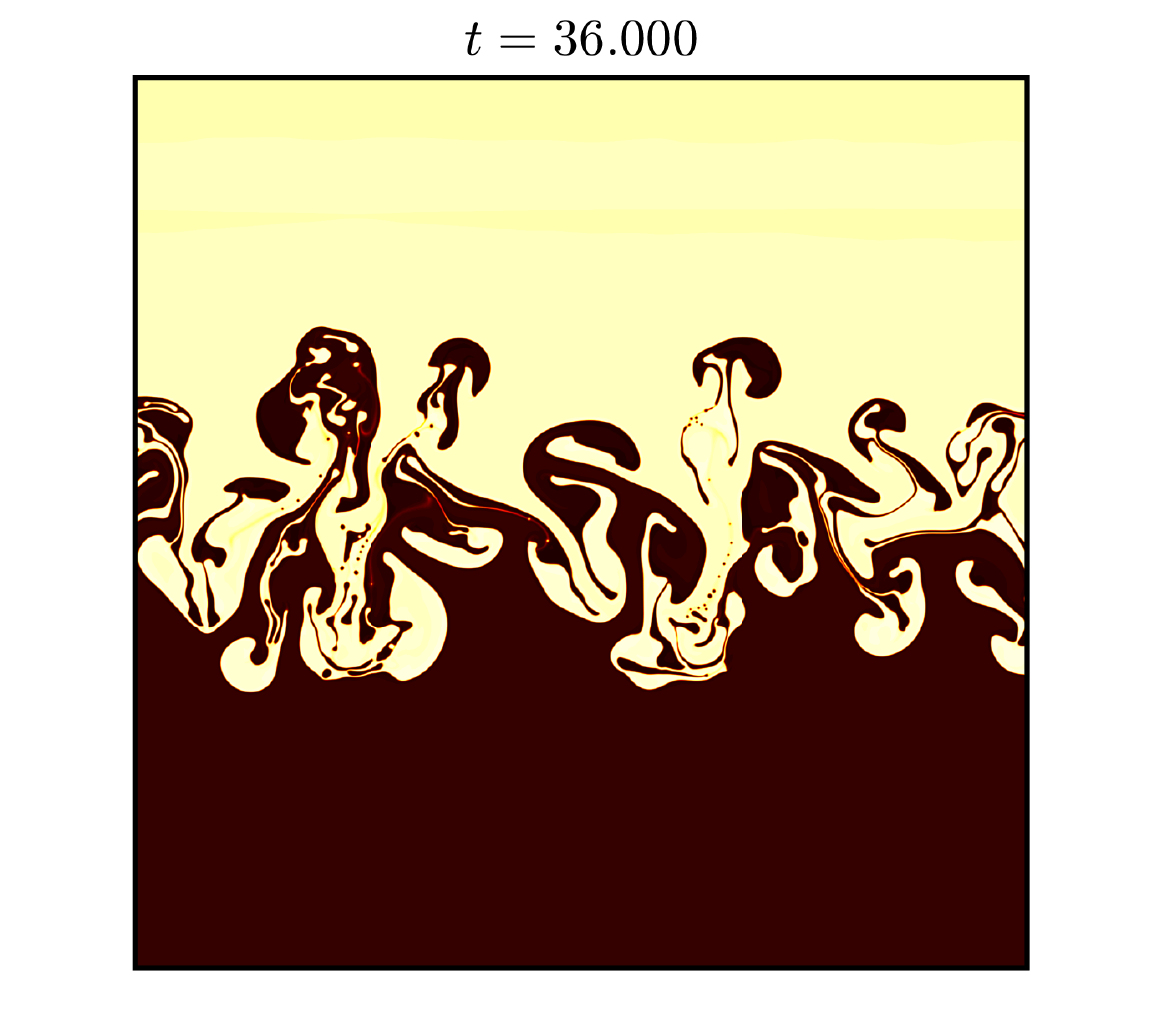}}
	\caption{Evolution of the interface profile for times close to the local maximum for the total interface energy; see also Fig.~\ref{Components of the pressure tensor}(d). For time $t=30.000$ the interface has only one component, while in the next few times the first topology changes take place leading to the first drops and large disconnected clusters.}
	\label{Critical point for interface energy}
\end{figure}

The graphs in Fig.~\ref{Components of the pressure tensor}(d) mark a transition to a new regime starting at times close to  $t = 30.000$, at which the variation of interface energy attains a local maximum. We argue that this transition indicates the moment when the interface becomes disconnected generating small drops and large disconnected clusters. This behavior is clearly seen in Fig.~\ref{Critical point for interface energy}, where we plot the density profiles at four different times. These figures show the process of how the interface becomes disconnected after $t = 30.000$. A similar phenomenon was observed in \cite{bertozzi2012diffuse} for the one-dimensional convective Cahn-Hilliard equation.

\subsection{Viscous dissipation and enstrophy}
\label{subsection enstrophy}

\begin{figure}[t]
	\centering
	\subfigure[]{\includegraphics[scale=0.45]{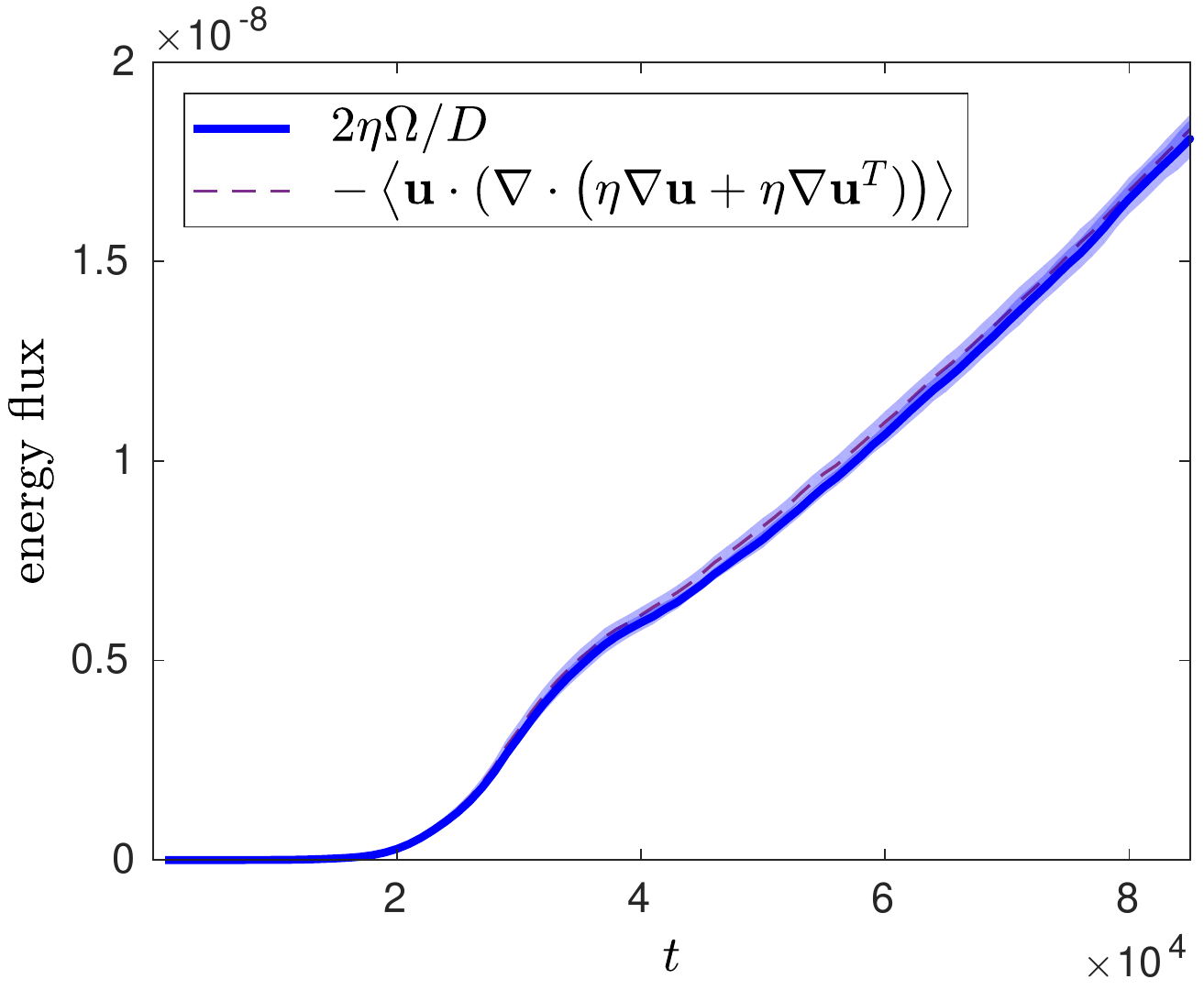}}\hspace{0.2cm}
	\subfigure[]{\includegraphics[scale=0.45]{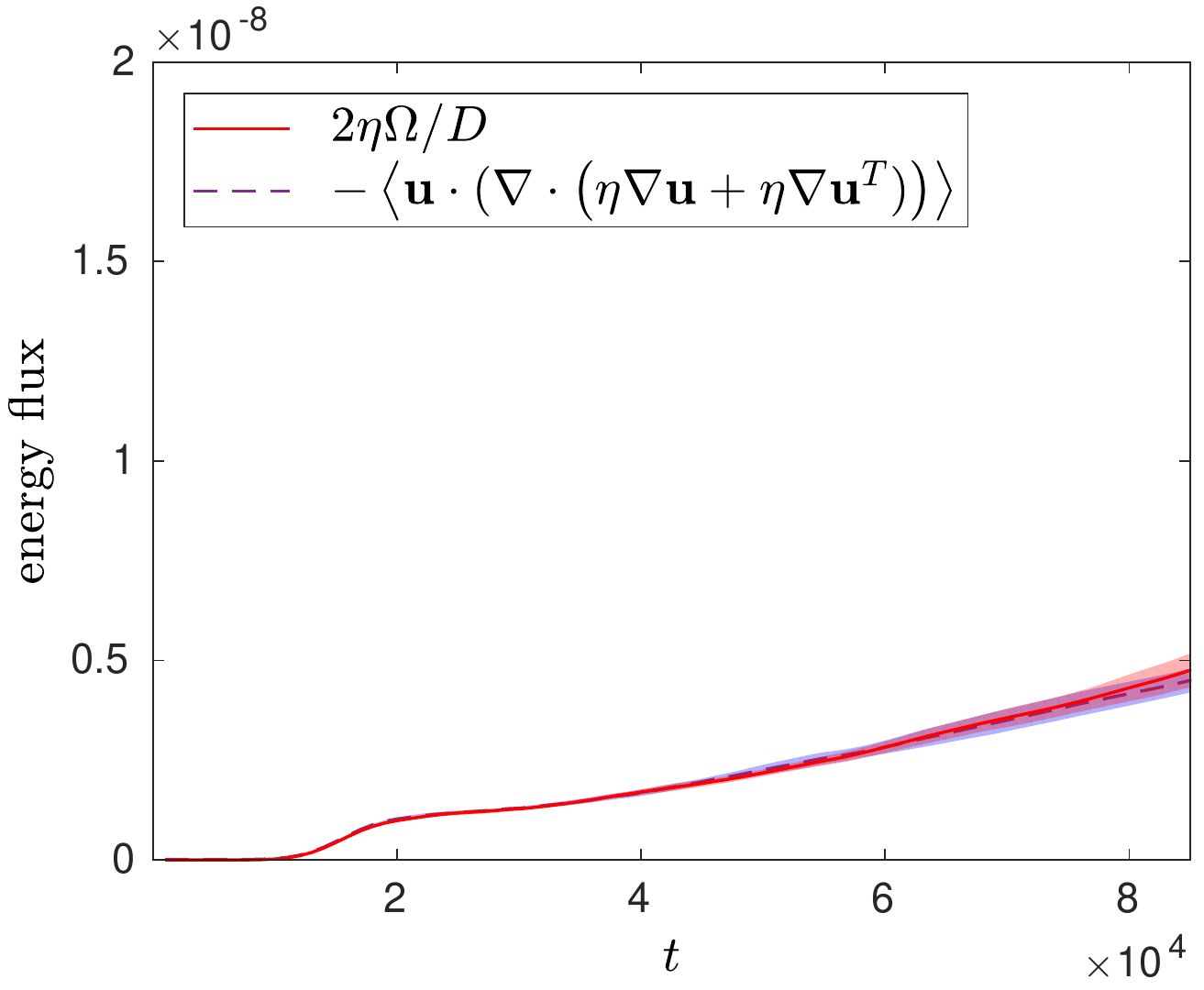}}\\
	\subfigure[]{\includegraphics[width=0.47\textwidth]{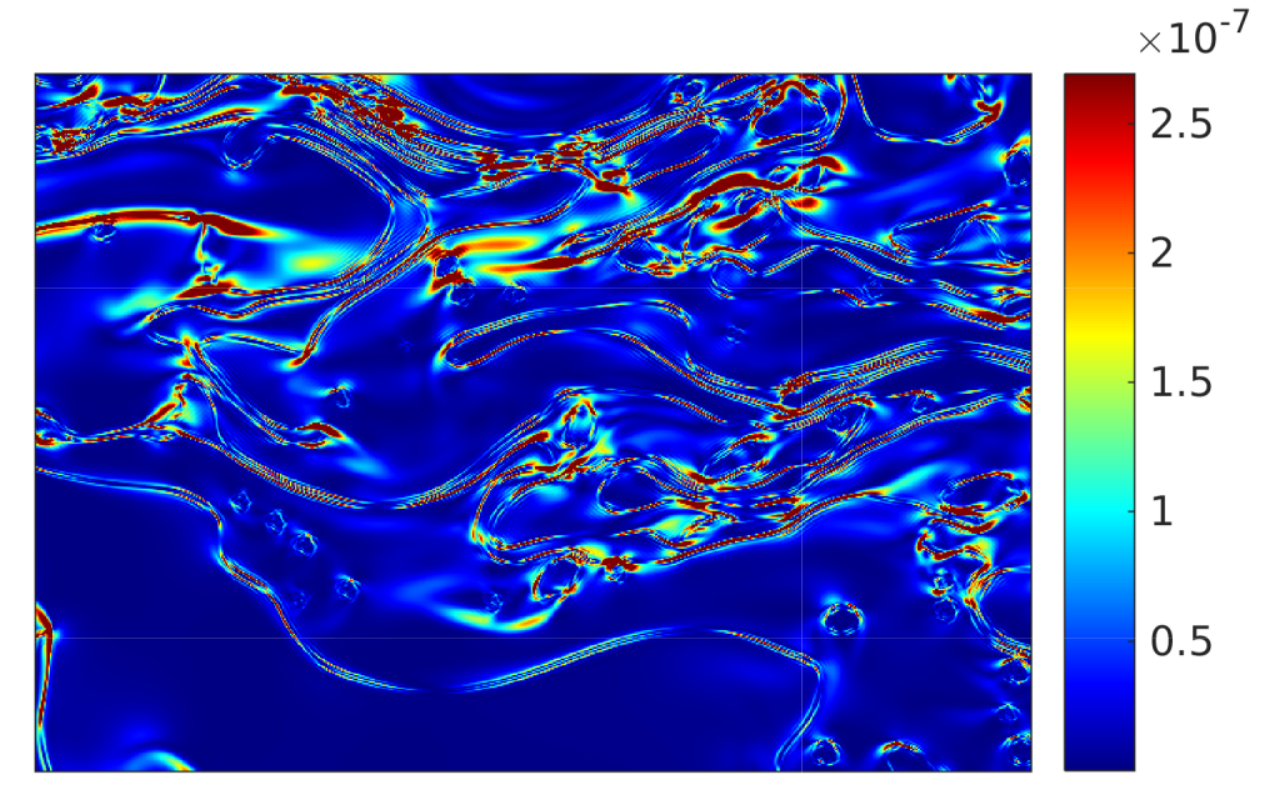}}\hspace{0.2cm}
	\subfigure[]{\includegraphics[width=0.47\textwidth]{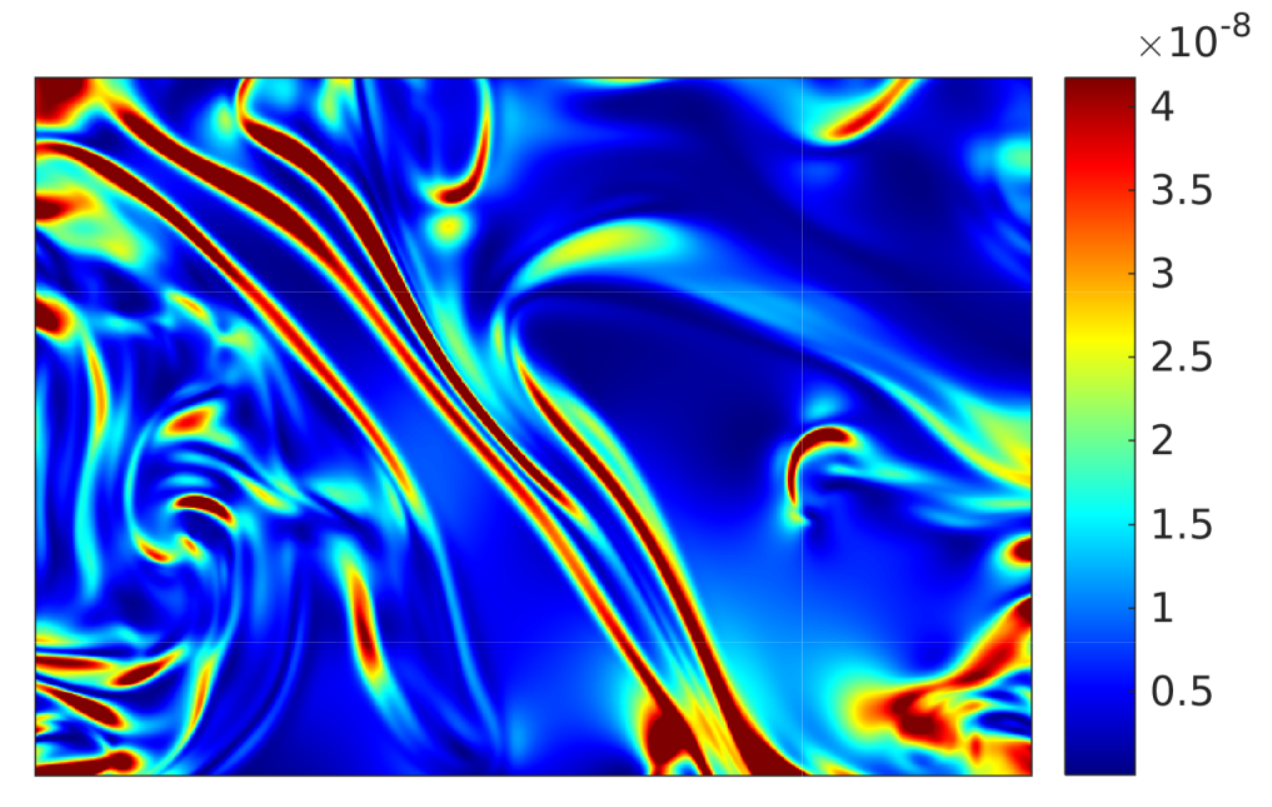}}
	\caption{Comparison between the evolution of the total enstrophy and the evolution of the dissipation function $\mathbf{u}\cdot (\nabla \cdot (\eta \nabla \mathbf{u} + \eta \nabla \mathbf{u}^T  ))$ for (a) immiscible and (b) miscible flows;
	shaded regions indicate standard deviations. Example of dissipation function for (c) immiscible and (d) miscible flows at $t=83.000$ corresponding to the final time in Figs.~\ref{fig1_novo} and~\ref{fig2_novo}.}
	\label{Enstrophy two}
\end{figure}

As follows from Figs.~\ref{Consistency checks and components of the  kinetic energy variation}(e, f), the most significant distinction between immiscible and miscible flows is related to the viscous dissipation function $\*u \cdot( \nabla \cdot \left( \eta \nabla \*u+\eta \nabla \*u^T)\right)$, whose mean value appears in the kinetic energy balance (\ref{Kinetic energy balance}). We now analyze the statistics of viscous dissipation,
exploring its connection with the enstrophy of the system. The enstrophy is defined as $\Omega=\frac{1}{2}\int \omega^2d\*x$,
where $\omega=\nabla \times \*u$ is a scalar vorticity of two-dimensional flow.
Neglecting the density variation, we have~\cite{koh1994vorticity}
\begin{equation}\label{Dissipation function}
\left\langle  \*u \cdot \left( \nabla \cdot \left( \eta \nabla \*u+\eta \nabla \*u^T \right)\right)\right\rangle= \eta \left\langle \*u \cdot \nabla^2 \*u \right\rangle=-\eta\left\langle |\nabla \times \*u|^2 \right\rangle=-\dfrac{2\eta}{D}\Omega.
\end{equation}
Relation (\ref{Dissipation function}) is verified for our LB simulations in Figs.~\ref{Enstrophy two}(a, b). These two figures also demonstrate a significant difference in the evolution of the viscous dissipation between immiscible and miscible flows.

It is apparent from Fig.~\ref{Enstrophy two}(a), that the values of the enstrophy are considerably larger for the immiscible flow. We now argue that this difference can be attributed to the flow in a small neighborhood of the interface. Figure~\ref{Enstrophy two}(c) shows the dissipation function for the immiscible flow; it corresponds to a small area of $667 \times 467$ lattice points marked by the rectangle in the center of Fig.~\ref{fig1_novo} and amplified in its right small panel corresponding to $t=83.000$. Visually, it is clear that a considerable part of high dissipation is concentrated near the interface. For comparison, we present the dissipation function for the miscible case in Fig.~\ref{Enstrophy two}(d), which corresponds to a small area from Fig.~\ref{fig2_novo}. In the miscible case, the dissipation is more dispersed and its amplitude is roughly half (notice the difference of color scales). 

 Part of the dissipation may have a numerical origin coming from spurious currents of the lattice Boltzmann method.  However, the experiments in~\cite{tavares2020immiscible,tavares2021} suggest that this numerical contribution is not  large enough to account for all the vorticity generated by the interface, once  typical values of spurious vorticity are same order or smaller than the values of the vorticity for the miscible case~\cite{tavares2021,kruger2017lattice}.

\vspace*{-5pt}

\section{Conclusion}

We have presented a high resolution study of the developed immiscible RT turbulence in 2D using the Shan-Chen multicomponent method. Through an appropriate choice of parameters, forcing scheme and initial configuration, we developed a LB model for simulating the immiscible and miscible RT systems in Boussinesq approximation. The simulation of the Shan-Chen multicomponent method in GPUs made it possible to collect a robust set of statistics, allowing direct verifications of phenomenological predictions for the RT turbulence~\cite{tavares2020immiscible}.
	
With the numerical results provided by the Shan-Chen model, we analyzed the energy budget of the RT turbulence. The verification of the energy balance shows the good accuracy of the solutions provided by the Shan-Chen model in solving the coupled Navier-Stokes and Cahn–Hilliard equations. In this analysis we found that the potential energy statistics are similar between immiscible and miscible flows. Significant differences were found in the statistics of kinetic energy. Analyzing the components of the kinetic energy variations, we found that the differences are associated with the momentum flux tensor and the viscous dissipation. We show that the flux due to the Korteweg stress tensor corresponds to the variation of the total length of the interface, calculated numerically by the Cauchy-Crofton formula. We show indications that the interface acts as a source of vorticity, which can explain a significant part of the difference in the viscous dissipation statistics verified between immiscible and miscible flows. The results for the miscible case are in line with the current studies about energy balance found in the scientific literature~\cite{zhao2020energy}. 
	
Analyzing the variation of the total interface energy, we verify the existence of a bifurcation point in the transition to turbulence. It is associated with the onset of topological changes in the interface between fluid phases. Before the critical point the interface has only one component, while after the critical point the interface is disconnected.  This process is characterized by the appearance of drops and isolated clusters evolving later to the emulsion-like mixed turbulent state. Regarding future works, the extension of the numerical procedures presented in this article for the three-dimensional immiscible Rayleigh-Taylor turbulence, which  is a more suitable configuration for experimental procedures, is an ongoing research. Such extension of the present GPU code can be a non-trivial task. This extension may also be  achieved by using some other lattice Boltzmann methods based on phase field models or the color gradient method, as in the works~\cite{leclaire2017generalized,de2019universal,hosseini2021lattice,liang2016lattice}, where the algorithms are employed in conjunction with MRT or central moments based algorithms.

\vskip6pt

\enlargethispage{20pt}


\dataccess{More information about the statistical analysis exposed in this article can be found in the thesis~\cite{tavares2021}. The files for the energy budget analysis and some
examples of velocity and density fields used in the article can be accessed in a data set via the following link: \url{https://doi.org/10.5061/dryad.4mw6m908p}.}\label{Data set}

\aucontribute{All authors contributed to the paper.}

\competing{The authors declare that they have no competing interests.}

\funding{AAM is supported by CNPq grants 303047/2018-6, 406431/2018-3 and FAPERJ Pensa Rio grant E-26/210.874/2014; and HST is supported by a doctoral scholarship from CAPES. This project has received funding from the European Research Council (ERC) under the European Union’s Horizon 2020 research and innovation programme (grant agreement No 882340.}

\ack{We thank Francesca Pelusi, from the University of Rome Tor Vergata, for the useful discussions on  technical  details  of  the  implementation  of  the  Shan-Chen  multicomponent  method.   We  also thank Sergio Pilotto and Daniel Lins de Albuquerque for their help in some aspects of the parallel implementation of the lattice Boltzmann method on GPUs. HST and AAM acknowledge the support from the ERC-ADG NewTURB project during their visits to the University of Rome – Tor Vergata.}


	\bibliographystyle{vancouver}
	
\bibliography{Bibliografia}

\begin{thebibliography}{10}

\bibitem{celani2009phase}
Celani A, Mazzino A, Muratore-Ginanneschi P, Vozella L.
\newblock Phase-field model for the {Rayleigh--Taylor} instability of
  immiscible fluids.
\newblock Journal of Fluid Mechanics. 2009;622:115--134.
\newblock (doi: \url{https://doi.org/10.1017/S0022112008005120}).

\bibitem{ramaprabhu2004experimental}
Ramaprabhu P, Andrews M.
\newblock Experimental investigation of {Rayleigh-Taylor} mixing at small
  Atwood numbers.
\newblock Journal of Fluid Mechanics. 2004;502:233.
\newblock (doi: \url{https://doi.org/10.1017/S0022112003007419}).

\bibitem{wilson1972excitatory}
Cowan JD, Wilson HR.
\newblock Excitatory and inhibitory interactions in localized populations of
  model neurons.
\newblock Biophysical Journal. 1972;12(1):1.
\newblock (doi: \url{https://doi.org/10.1016/S0006-3495(72)86068-5}).

\bibitem{huang2007rayleigh}
Huang Z, De~Luca A, Atherton TJ, Bird M, Rosenblatt C, Carles P.
\newblock Rayleigh-Taylor instability experiments with precise and arbitrary
  control of the initial interface shape.
\newblock Physical review letters. 2007;99(20):204502.
\newblock (doi: \url{https://doi.org/10.1103/PhysRevLett.99.204502}).

\bibitem{waddell2001experimental}
Waddell J, Niederhaus C, Jacobs JW.
\newblock Experimental study of Rayleigh--Taylor instability: low Atwood number
  liquid systems with single-mode initial perturbations.
\newblock Physics of Fluids. 2001;13(5):1263--1273.
\newblock (doi: \url{https://doi.org/10.1063/1.1359762}).

\bibitem{chertkov2003phenomenology}
Chertkov M.
\newblock Phenomenology of {Rayleigh-Taylor} turbulence.
\newblock Physical Review Letters. 2003;91(11):115001.
\newblock (doi: \url{https://doi.org/10.1103/PhysRevLett.91.115001}).

\bibitem{biferale2010high}
Biferale L, Mantovani F, Sbragaglia M, Scagliarini A, Toschi F, Tripiccione R.
\newblock High resolution numerical study of Rayleigh--Taylor turbulence using
  a thermal lattice Boltzmann scheme.
\newblock Physics of Fluids. 2010;22(11):115112.
\newblock Available from: \url{https://doi.org/10.1063/1.3517295}.

\bibitem{celani2006rayleigh}
Celani A, Mazzino A, Vozella L.
\newblock {Rayleigh-Taylor} turbulence in two dimensions.
\newblock Physical Review Letters. 2006;96(13):134504.

\bibitem{boffetta2017incompressible}
Boffetta G, Mazzino A.
\newblock Incompressible {Rayleigh--Taylor} turbulence.
\newblock Annual Review of Fluid Mechanics. 2017;49:119--143.
\newblock (doi: \url{https://doi.org/10.1146/annurev-fluid-010816-060111}).

\bibitem{biferale2018rayleigh}
Biferale L, Boffetta G, Mailybaev AA, Scagliarini A.
\newblock {Rayleigh-Taylor turbulence with singular nonuniform initial
  conditions}.
\newblock Physical Review Fluids. 2018;3(9):092601.
\newblock (doi: \url{https://doi.org/10.1103/PhysRevFluids.3.092601}).

\bibitem{zhao2020energy}
Zhao D, Aluie H.
\newblock Energy cascades in Rayleigh-Taylor turbulence.
\newblock arXiv preprint arXiv:200604301. 2020.

\bibitem{zhou2017rayleigh}
Zhou Y.
\newblock Rayleigh--Taylor and Richtmyer--Meshkov instability induced flow,
  turbulence, and mixing. II.
\newblock Physics Reports. 2017;723:1--160.
\newblock (\mbox{doi}: \url{https://doi.org/10.1016/j.physrep.2017.07.008}).

\bibitem{young2006surface}
Young YN, Ham F.
\newblock Surface tension in incompressible {Rayleigh--Taylor} mixing flow.
\newblock Journal of Turbulence. 2006;(7):N71.
\newblock (doi: \url{https://doi.org/10.1080/14685240600809979}).

\bibitem{brackbill1992continuum}
Brackbill JU, Kothe DB, Zemach C.
\newblock A continuum method for modeling surface tension.
\newblock Journal of computational physics. 1992;100(2):335--354.
\newblock (doi: \url{https://doi.org/10.1016/0021-9991(92)90240-Y}).

\bibitem{carles2006rayleigh}
Carles P, Huang Z, Carbone G, Rosenblatt C.
\newblock Rayleigh-Taylor instability for immiscible fluids of arbitrary
  viscosities: A magnetic levitation investigation and theoretical model.
\newblock Physical Review Letters. 2006;96(10):104501.
\newblock (doi: \url{https://doi.org/10.1103/PhysRevLett.96.104501}).

\bibitem{livescu2004compressibility}
Livescu D.
\newblock Compressibility effects on the Rayleigh--Taylor instability growth
  between immiscible fluids.
\newblock Physics of fluids. 2004;16(1):118--127.
\newblock (doi: \url{https://doi.org/10.1063/1.1630800}).

\bibitem{liang2019direct}
Liang H, Hu X, Huang X, Xu J.
\newblock Direct numerical simulations of multi-mode immiscible
  {Rayleigh-Taylor} instability with high Reynolds numbers.
\newblock Physics of Fluids. 2019;31(11):112104.
\newblock (doi: \url{https://doi.org/10.1063/1.5127888}).

\bibitem{huang2020late}
Huang H, Xia Z, Liang H, Zong Y, Xu J.
\newblock Late-time description of immiscible Rayleigh-Taylor instability: A
  lattice Boltzmann study.
\newblock arXiv preprint arXiv:200914655. 2020.

\bibitem{hosseini2021lattice}
Hosseini SA, Safari H, Thevenin D.
\newblock Lattice Boltzmann Solver for Multiphase Flows: Application to High
  Weber and Reynolds Numbers.
\newblock Entropy. 2021;23(2):166.
\newblock (doi: \url{ https://doi.org/10.3390/e23020166}).

\bibitem{liang2016lattice}
Liang H, Li Q, Shi B, Chai Z.
\newblock Lattice Boltzmann simulation of three-dimensional Rayleigh-Taylor
  instability.
\newblock Physical Review E. 2016;93(3):033113.
\newblock (doi: \url{https://doi.org/10.1103/PhysRevE.93.033113}).

\bibitem{tavares2020immiscible}
Tavares HS, Biferale L, Sbragaglia M, Mailybaev AA.
\newblock Immiscible Rayleigh-Taylor turbulence using mesoscopic lattice
  Boltzmann algorithms.
\newblock Physical Review Fluids. 2021;6(5):054606.
\newblock (doi: \url{https://doi.org/10.1103/PhysRevFluids.6.054606}).

\bibitem{liang2014phase}
Liang H, Shi B, Guo Z, Chai Z.
\newblock Phase-field-based multiple-relaxation-time lattice Boltzmann model
  for incompressible multiphase flows.
\newblock Physical Review E. 2014;89(5):053320.
\newblock (doi: \url{https://doi.org/10.1103/PhysRevE.89.053320}).

\bibitem{wang2019brief}
Wang H, Yuan X, Liang H, Chai Z, Shi B.
\newblock A brief review of the phase-field-based lattice Boltzmann method for
  multiphase flows.
\newblock Capillarity. 2019;2(3):33--52.
\newblock (doi: \url{https://doi.org/10.26804/capi.2019.03.01}).

\bibitem{chertkov2005effects}
Chertkov M, Kolokolov I, Lebedev V.
\newblock Effects of surface tension on immiscible {Rayleigh-Taylor}
  turbulence.
\newblock Physical Review E. 2005;71(5):055301.

\bibitem{kruger2017lattice}
Kr{\"u}ger T, Kusumaatmaja H, Kuzmin A, Shardt O, Silva G, Viggen EM.
\newblock The lattice {Boltzmann} method.
\newblock Springer International Publishing. 2017;10:978--3.
\newblock (doi: \url{https://doi.org/10.1007/978-3-319-44649-3}).

\bibitem{succi2018lattice}
Succi S.
\newblock The lattice Boltzmann equation: for complex states of flowing matter.
\newblock Oxford University Press; 2018.
\newblock (doi: \url{https://doi.org/10.1093/oso/9780199592357.001.0001}).

\bibitem{falcucci2007lattice}
Falcucci G, Bella G, Chiatti G, Chibbaro S, Sbragaglia M, Succi S, et~al.
\newblock Lattice Boltzmann models with mid-range interactions.
\newblock Communications in computational physics. 2007;2(6):1071--1084.

\bibitem{scarbolo2013unified}
Scarbolo L, Molin D, Perlekar P, Sbragaglia M, Soldati A, Toschi F.
\newblock Unified framework for a side-by-side comparison of different
  multicomponent algorithms: Lattice {Boltzmann} vs. phase field model.
\newblock Journal of Computational Physics. 2013;234:263--279.
\newblock (doi: \url{https://doi.org/10.1016/j.jcp.2012.09.029}).

\bibitem{chikatamarla2015entropic}
Chikatamarla S, Karlin I, et~al.
\newblock Entropic lattice Boltzmann method for multiphase flows.
\newblock Physical review letters. 2015;114(17):174502.
\newblock (doi: \url{https://doi.org/10.1103/PhysRevLett.114.174502}).

\bibitem{leclaire2017generalized}
Leclaire S, Parmigiani A, Malaspinas O, Chopard B, Latt J.
\newblock Generalized three-dimensional lattice Boltzmann color-gradient method
  for immiscible two-phase pore-scale imbibition and drainage in porous media.
\newblock Physical Review E. 2017;95(3):033306.
\newblock (doi: \url{ https://doi.org/10.1103/PhysRevE.95.033306}).

\bibitem{geier2015conservative}
Geier M, Fakhari A, Lee T.
\newblock Conservative phase-field lattice Boltzmann model for interface
  tracking equation.
\newblock Physical Review E. 2015;91(6):063309.
\newblock (doi: \url{https://doi.org/10.1103/PhysRevE.91.063309}).

\bibitem{de2019universal}
De~Rosis A, Huang R, Coreixas C.
\newblock Universal formulation of central-moments-based lattice Boltzmann
  method with external forcing for the simulation of multiphysics phenomena.
\newblock Physics of Fluids. 2019;31(11):117102.
\newblock (doi: \url{https://doi.org/10.1063/1.5124719}).

\bibitem{sbragaglia2007generalized}
Sbragaglia M, Benzi R, Biferale L, Succi S, Sugiyama K, Toschi F.
\newblock Generalized lattice {Boltzmann} method with multirange
  pseudopotential.
\newblock Physical Review E. 2007;75(2):026702.
\newblock (doi: \url{https://doi.org/10.1103/PhysRevE.75.026702}).

\bibitem{connington2012review}
Connington K, Lee T.
\newblock A review of spurious currents in the lattice {Boltzmann} method for
  multiphase flows.
\newblock Journal of Mechanical Science and Technology. 2012;26(12):3857--3863.
\newblock (doi: \url{https://doi.org/10.1007/s12206-012-1011-5}).

\bibitem{blanchette2009energy}
Blanchette F, Lei Y.
\newblock Energy considerations for multiphase fluids with variable density and
  surface tension.
\newblock SIAM review. 2009;51(2):423--431.
\newblock (doi: \url{https://doi.org/10.1137/070694880}).

\bibitem{desai2009dynamics}
Desai RC, Kapral R.
\newblock Dynamics of Self-organized and Self-assembled Structures.
\newblock Cambridge University Press; 2009.
\newblock (doi: \url{https://doi.org/10.1017/CBO9780511609725}).

\bibitem{guo2015thermodynamically}
Guo Z, Lin P.
\newblock A thermodynamically consistent phase-field model for two-phase flows
  with thermocapillary effects.
\newblock Journal of Fluid Mechanics. 2015;766:226--271.
\newblock (doi: \url{https://doi.org/10.1017/jfm.2014.696}).

\bibitem{liu2003phase}
Liu C, Shen J.
\newblock A phase field model for the mixture of two incompressible fluids and
  its approximation by a Fourier-spectral method.
\newblock Physica D: Nonlinear Phenomena. 2003;179(3-4):211--228.
\newblock (doi: \url{https://doi.org/10.1016/S0167-2789(03)00030-7}).

\bibitem{terrington2020generation}
Terrington S, Hourigan K, Thompson M.
\newblock The generation and conservation of vorticity: deforming interfaces
  and boundaries in two-dimensional flows.
\newblock Journal of Fluid Mechanics. 2020;890.
\newblock (doi: \url{https://doi.org/10.1017/jfm.2020.128}).

\bibitem{brons2014vorticity}
Br{\o}ns M, Thompson MC, Leweke T, Hourigan K.
\newblock Vorticity generation and conservation for two-dimensional interfaces
  and boundaries.
\newblock Journal of Fluid Mechanics. 2014;758:63--93.
\newblock (doi: \url{https://doi.org/10.1017/jfm.2014.520}).

\bibitem{anderson1998diffuse}
Anderson DM, McFadden GB, Wheeler AA.
\newblock Diffuse-interface methods in fluid mechanics.
\newblock Annual review of fluid mechanics. 1998;30(1):139--165.
\newblock (doi: \url{https://doi.org/10.1146/annurev.fluid.30.1.139}).

\bibitem{kullmer2018transition}
K{\"u}llmer K, Kr{\"a}mer A, Joppich W, Reith D, Foysi H.
\newblock Transition point prediction in a multicomponent lattice Boltzmann
  model: Forcing scheme dependencies.
\newblock Physical Review E. 2018;97(2):023313.
\newblock (doi: \url{https://doi.org/10.1103/PhysRevE.97.023313}).

\bibitem{benzi2009mesoscopic}
Benzi R, Sbragaglia M, Succi S, Bernaschi M, Chibbaro S.
\newblock Mesoscopic lattice {Boltzmann} modeling of soft-glassy systems:
  theory and simulations.
\newblock The Journal of Chemical Physics. 2009;131(10):104903.
\newblock (doi: \url{https://doi.org/10.1063/1.3216105}).

\bibitem{joseph1996non}
Joseph DD, Huang A, Hu H.
\newblock Non-solenoidal velocity effects and {Korteweg} stresses in simple
  mixtures of incompressible liquids.
\newblock Physica D: Nonlinear Phenomena. 1996;97(1-3):104--125.
\newblock (doi: \url{https://doi.org/10.1016/0167-2789(96)00097-8}).

\bibitem{joseph2010fluid}
Joseph DD.
\newblock Fluid Dynamics of Mixtures of Incompressible Miscible Liquids.
\newblock In: Applied and Numerical Partial Differential Equations. Springer;
  2010. p. 127--145.
\newblock (doi: \url{https://doi.org/10.1007/978-90-481-3239-3_10}).

\bibitem{li2020multiscale}
Li J.
\newblock Multiscale and multiphysics flow simulations of using the {Boltzmann}
  equation.
\newblock Springer; 2020.

\bibitem{shan1996diffusion}
Shan X, Doolen G.
\newblock Diffusion in a multicomponent lattice Boltzmann equation model.
\newblock Physical Review E. 1996;54(4):3614.
\newblock (doi: \url{https://doi.org/10.1103/PhysRevE.54.3614}).

\bibitem{bernaschi2009graphics}
Bernaschi M, Rossi L, Benzi R, Sbragaglia M, Succi S.
\newblock Graphics processing unit implementation of lattice Boltzmann models
  for flowing soft systems.
\newblock Physical Review E. 2009;80(6):066707.
\newblock (doi: \url{https://doi.org/10.1103/PhysRevE.80.066707}).

\bibitem{bernaschi2017gpu}
Bernaschi M, Lulli M, Sbragaglia M.
\newblock GPU based detection of topological changes in Voronoi diagrams.
\newblock Computer Physics Communications. 2017;213:19--28.
\newblock (doi: \url{https://doi.org/10.1016/j.cpc.2016.11.005}).

\bibitem{pelusi2019impact}
Pelusi F, Sbragaglia M, Scagliarini A, Lulli M, Bernaschi M, Succi S.
\newblock On the impact of controlled wall roughness shape on the flow of a
  soft material.
\newblock EPL (Europhysics Letters). 2019;127(3):34005.
\newblock (doi: \url{https://doi.org/10.1209/0295-5075/127/34005}).

\bibitem{tavares2021}
Tavares HS.
\newblock Lattice Boltzmann modelling for immiscible Rayleigh-Taylor turbulence
  [Ph.D. thesis].
\newblock Institute for Pure and Applied Mathematics (IMPA); 2021.
\newblock Available from:
  \url{https://impa.br/wp-content/uploads/2021/03/tese_dout_Hugo-Saraiva-Tavares.pdf}.

\bibitem{chella1996mixing}
Chella R, Vi{\~n}als J.
\newblock Mixing of a two-phase fluid by cavity flow.
\newblock Physical Review E. 1996;53(4):3832.
\newblock (doi: \url{https://doi.org/10.1103/PhysRevE.53.3832}).

\bibitem{scarbolo2015coalescence}
Scarbolo L, Bianco F, Soldati A.
\newblock Coalescence and breakup of large droplets in turbulent channel flow.
\newblock Physics of Fluids. 2015;27(7):073302.
\newblock (doi: \url{https://doi.org/10.1063/1.4923424}).

\bibitem{goldman2005curvature}
Goldman R.
\newblock Curvature formulas for implicit curves and surfaces.
\newblock Computer Aided Geometric Design. 2005;22(7):632--658.
\newblock (doi: \url{https://doi.org/10.1016/j.cagd.2005.06.005}).

\bibitem{sevcovic2001evolution}
Sevcovic D, Mikula K.
\newblock Evolution of plane curves driven by a nonlinear function of curvature
  and anisotropy.
\newblock SIAM Journal on Applied Mathematics. 2001;61(5):1473--1501.
\newblock (doi: \url{https://doi.org/10.1137/S0036139999359288}).

\bibitem{do2016differential}
Do~Carmo MP.
\newblock Differential geometry of curves and surfaces: revised and updated
  second edition.
\newblock Courier Dover Publications; 2016.

\bibitem{legland2007computation}
Legland D, Ki{\^e}u K, Devaux MF.
\newblock Computation of {Minkowski} measures on {2D} and {3D} binary images.
\newblock Image Analysis \& Stereology. 2007;26(2):83--92.
\newblock (doi: \url{https://doi.org/10.5566/ias.v26.p83-92}).

\bibitem{bertozzi2012diffuse}
Bertozzi A, Kolokolnikov T, Liu W.
\newblock Diffuse interface surface tension models in an expanding flow.
\newblock Communications in Mathematical Sciences. 2012;10(1):387--418.
\newblock (doi: \url{https://dx.doi.org/10.4310/CMS.2012.v10.n1.a16}).

\bibitem{koh1994vorticity}
Koh YM.
\newblock Vorticity and viscous dissipation in an incompressible flow.
\newblock KSME Journal. 1994;8(1):35--42.
\newblock (doi: \url{https://doi.org/10.1007/BF02953241}).

\end{thebibliography}

\end{document}